\documentclass[twocolumn]{aastex701}

\newcommand{\pc}[2]{\parbox[t]{#1}{#2\strut}}
\newcommand{\pcl}[2]{\parbox[t]{#1}{\raggedright #2\strut}}

\usepackage{amsmath,amssymb,amsfonts,amsthm}
\usepackage{graphicx}
\usepackage{bm}
\makeatletter
\newcommand\algorithmname{Algorithm}
\newcounter{algorithm}
\def\fps@algorithm{tbp}
\def\ftype@algorithm{4}
\def\ext@algorithm{loa}
\def\fnum@algorithm{\algorithmname~\thealgorithm}
\newenvironment{algorithm}{\@float{algorithm}}{\end@float}
\newenvironment{algorithm*}{\@dblfloat{algorithm}}{\end@dblfloat}
\makeatother
\usepackage{algorithmic}
\usepackage{booktabs}
\usepackage{xcolor}
\definecolor{aspred}{RGB}{180,30,45}

\newcommand{\vect}[1]{\bm{#1}}
\newcommand{\mat}[1]{\mathbf{#1}}

\newcommand{\reals}{\mathbb{R}}

\newcommand{\mss}{\ensuremath{\mathrm{m\,s^{-2}}}}  

\DeclareMathOperator*{\argmin}{arg\,min}

\shorttitle{NestyNet. III. Symbolic Regression from Analytic Surrogates}
\shortauthors{Ibata et al.}

\begin{document}

\title{NestyNet. III. Symbolic Regression from Analytic Neural Surrogates}

\author[0000-0002-3292-9709]{Rodrigo Ibata}
\affiliation{Universit\'e de Strasbourg, CNRS, Observatoire astronomique de Strasbourg, UMR 7550, F-67000 Strasbourg, France}
\email[show]{rodrigo.ibata@astro.unistra.fr}

\author[0000-0001-8392-3836]{Wassim Tenachi}
\affiliation{Mila - Quebec Artificial Intelligence Institute}
\affiliation{D\'epartement de physique, Universit\'e de Montr\'eal}
\affiliation{Ciela - Montreal Institute for Astrophysical Data Analysis and Machine Learning}
\email{wassim.tenachi@umontreal.ca}

\author[0000-0002-8788-8174]{Foivos Diakogiannis}
\affiliation{Technology, Commonwealth Scientific and Industrial Research Organisation (CSIRO), Kensington, WA 6151, Australia}
\email{Foivos.Diakogiannis@data61.csiro.au}

\author[0000-0003-1559-1053]{Neil Ibata}
\affiliation{Department of Human Evolutionary Biology, Harvard University, Cambridge, MA, USA}
\email{neilibata@fas.harvard.edu}

\author[0009-0008-7455-1880]{Anirudh Shankar}
\affiliation{Universit\'e de Strasbourg, CNRS, Observatoire astronomique de Strasbourg, UMR 7550, F-67000 Strasbourg, France}
\email{anirudh.shankar@astro.unistra.fr}

\begin{abstract}
Many physical laws are simple only after the right representation,
decomposition or internal coordinate has been found, but discovering that
structure from data is combinatorially hard. This task is symbolic
regression (SR), the search for closed-form expressions that fit data
without assuming a fixed model class. Here we present NestyNet-SR.  
A neural surrogate with analytic derivatives
is used to detect separability, recursively reducing multivariate problems
to simpler neural atoms.  These atoms are distilled into closed form
by a tiered symbolic-search stack, whose final tier is a novel factorized
symbolic search that separates structure from calibration.  Composing
candidate internal coordinates freely, it scores each coordinate by how
well calibrated functions of it (e.g., polynomials, power laws, sinusoids)
fit the data, so the constants of those calibrated maps, however deeply
nested in the final expression, are fitted rather than searched.  The method supports multi-dataset regression,
automated feature discovery, and dimensional-analysis pruning.  On the
SRBench AI~Feynman benchmark, NestyNet-SR achieves exact symbolic recovery
of all 120 noiseless equations, the first such result, and under noise a
statistical audit certifies which structures survive.  As a real-data
vignette, given only the separate mass-model components of SPARC-survey galaxies,
the algorithm discovers the baryonic acceleration coordinate, reproduces
the established mass-to-light and acceleration scales and the non-unique
form of the radial acceleration relation, and adds
held-out-galaxy generalization, a calibrated symmetry abstention,
and a posterior for the local slope of the law.  Analytic derivatives thus
provide a practical route from
neural surrogates to interpretable closed-form empirical laws.
\end{abstract}


\section{Introduction}
\label{sec:introduction}

Is it hubris to expect that all interesting laws of nature, at all physical
scales, governing all physical or even abstract structures, are
discoverable through direct human intuition?  Or might machine-assisted
exploration help us survey a much vaster range of possibilities than our
unaided minds, to uncover empirical relations that can be verified on
held-out data, and yet still provide accurate predictions in new regimes
beyond the original data range?  Such empirical laws might then be studied
to yield a theoretical understanding that could lead to the discovery of new
underlying natural laws.  This is one of the motivations behind symbolic
regression (SR), the search for functions that fit data without assuming a
prescribed model class.

The difficulty with SR is that the space of candidate expressions is
combinatorially vast, so practical methods must exploit strong inductive
biases to cut the search down to a manageable size.  Separability is an
especially powerful such bias, since many scientific models decompose into
additive, multiplicative, or compound-coordinate structure, and recognizing
this decomposition turns an $n$-variable problem into a sequence of
simpler, lower-dimensional subproblems.

Another bias, which one can perhaps view as the foundational assumption of
physics, is that interesting natural laws are simple when interpreted
correctly, for instance in an appropriate coordinate system.  This principle
motivates a search not only over expressions but also over
\emph{representations}.  A relationship that appears intractable in its raw
form may become trivially separable after a change of output variable, a
reparameterization of the inputs, or the recognition of a compound coordinate.

This decomposition viewpoint was crystallized by the AI~Feynman
team~\citep{Udrescu2020,Udrescu2020b}, which showed that symmetry,
separability, and graph modularity can recursively reduce
symbolic regression problems to simpler subproblems.  We adopt that
perspective directly in the present work.  Our first contribution is to
make decomposition motifs reliable in practice using analytic derivatives,
and to extend the search to richer internal coordinates than the original
motif set.

Several other lines of work attack the same problem.  Genetic-programming (GP)
systems~\citep{Koza1992,Schmidt2009} evolve populations of expression trees
through mutation and crossover.  The Eureqa system~\citep{Schmidt2009}
famously rediscovered Hamiltonians, Lagrangians, and conservation laws from
pendulum data, and modern variants
such as PySR~\citep{PySR2023} couple GP with gradient-based coefficient
optimization on a Pareto frontier.  Reinforcement-learning
approaches~\citep{Petersen2021,Tenachi2023PhySO} cast token generation as a
sequential decision process built on REINFORCE-style
estimators~\citep{Williams1992}, and dimensional-analysis constraints can be directly
embedded into the prior~\citep{Tenachi2023PhySO}.  Neural-guided
methods~\citep{Cranmer2020} extract symbolic expressions from learned
graph-network representations, and mixed-integer nonlinear optimization~\citep{Austel2020}
or Monte Carlo tree search~\citep{Sun2023MCTS} navigate the expression-tree
space with varying degrees of guidance.  A common feature is that these
approaches entangle structural search with coefficient fitting.  Constants
are proposed as discrete tokens or optimized one finished candidate at a
time, so a correct inner structure earns no credit until its constants have
also been found.  Some recent work~\citep{Li2023TJSL} frames the problem as
skeleton recovery followed by coefficient optimization, a factorization the
present work pushes considerably further.

The challenge, as we found early in this work, is that reliable separability
tests require accurate function values and accurate mixed second
derivatives, whereas standard neural surrogates are generally insufficiently
accurate for this purpose, leading to both false positives and false
negatives in structure detection.  Solving this issue was one of the motivations for
the development and design of the NestyNet algorithm~\citep[hereafter Paper~I]{NestyNet2026a},
which is able to provide very accurate numerical
representations of functions and their derivatives.  Likewise, converting a
trained surrogate into a compact closed-form expression poses a combinatorial
challenge of its own, one that we quantify in the next section.

Here we introduce NestyNet-SR, a symbolic regression framework that uses
accurate NestyNet surrogate models as differentiable empirical
representations from which symbolic structure can be extracted.
Our second contribution is the factorized symbolic search (FSS) layer at its
core, motivated by the observation that many
scientific expressions are difficult not because their final outer form is
complicated, but because the correct internal coordinate is hidden.  Once a
carrier such as $x_0x_1$, $x_0/x_1$, $x_0-x_1$, or a low-depth composition is
found, the remaining map from carrier to target is often polynomial,
rational, trigonometric, exponential, or a low-dimensional calibrated
variant thereof.  NestyNet-SR therefore searches over carriers and scores
entire families of outer renderings at once.  To organize that search it
\emph{fingerprints} the residual of each partially fitted candidate (that is,
it reduces the residual to a compact, hash-like signature), groups
candidates with similar failure modes into residual basins, and reuses them
as seeds for mutation and repair.


The remainder of this paper is organized as follows.
Section~\ref{sec:design} lays out the overall two-stage design, in which a
surrogate-decomposition stage (Stage~A) reduces the problem to simple
neural building blocks and a symbolic-rewrite stage (Stage~B) distills
them into closed form, together with the abstract-syntax-tree (AST)
representation that underpins all our symbolic expressions.  Section~\ref{sec:stageA} describes the
Stage~A separability detection algorithms.  We turn to the core analytical discovery machinery in
\S\ref{sec:factorized-search}, which presents the factorized symbolic-search
layer.  Section~\ref{sec:gs} then develops a
generalized-symmetry layer that unifies the Stage~A coordinate detectors
under a single determining operator and can supply internal coordinates to
the factorized search, while \S\ref{sec:stageB} describes the
complementary Stage~B rewrite rules.  Section~\ref{sec:features} covers the automated
feature discovery that guides Stage~B, and \S\ref{sec:ytransforms}
discusses the Y-transforms mechanism.
Section~\ref{sec:units} discusses dimensional analysis and units
enforcement.  We present benchmarks in
\S\ref{sec:benchmarks}, both of the full pipeline and of the factorized
search in isolation, together with a real-data vignette recovering the
baryonic acceleration coordinate from SPARC galaxies, and conclude with
a summary and discussion of future directions in
\S\ref{sec:conclusions}.

\section{The Symbolic Regression Problem}

Symbolic regression, stated formally, seeks a vector-valued function
$\vect{f}^* \in \mathcal{F}$ mapping inputs $\vect{x}\in\reals^{n}$ to
outputs $\vect{y}\in\reals^{M}$ that minimizes reconstruction error while
maintaining low complexity,
\begin{equation}
\vect{f}^* = \argmin_{\vect{f} \in \mathcal{F}} \sum_{i=1}^{N} \|\vect{y}_i - \vect{f}(\vect{x}_i)\|^2 + \lambda \, C(\vect{f})
\label{eq:sr_objective}
\end{equation}
where $\mathcal{F}$ is the space of mathematical expressions (typically
represented as expression trees, one per output component) and $C(\vect{f})$
measures complexity (tree depth, number of operations, and so forth).  In
the scalar case $M=1$ each expression tree maps $\reals^n\to\reals$. For
vector-valued problems ($M>1$), the outputs may share structure or be
discovered independently.  The challenge is that $\mathcal{F}$ is enormous.
On the AI~Feynman benchmark, the longest target expression contains 40 nodes
in a standard binary expression-tree representation.  Labeling even that
fixed tree from the benchmark's operators and variables already yields
$\approx 1.4 \times 10^{32}$ candidates, and the space is vastly larger if
the tree structure itself is allowed to vary.  Symbolic regression in this
general form has in fact been proven
NP-hard~\citep{Virgolin2022NPhard}.  Intelligent decomposition is therefore
essential.

\subsection{Separability: The Key to Tractability}

Physical equations very often exhibit \emph{separability} in a broad sense.
Their dependence can be organized into lower-dimensional modules, as in the
clean additive and multiplicative splits $f(x)=g(u)+h(v)$ and
$f(x)=g(u)\,h(v)$ for disjoint variable groups $u$ and $v$, or into a
simpler internal coordinate $z(x)$.  Our search treats separability as a
spectrum of structural simplifications.
We allow approximate and overlapping decompositions, so a variable may
appear in more than one factor when that better reflects the governing law,
and we search for mixed cases in which an internal coordinate renders the
remaining dependence effectively univariate or of low \emph{arity}
(depending on few input arguments).  For instance,
$f(x_1,x_2,x_3)=x_1\,g(x_1x_2/x_3)$ is not classically separable, but it
becomes tractable once the internal coordinate is identified.  Recursive
application of this principle reduces many complicated multivariate
expressions to simple low-dimensional functions that the pipeline then
rewrites analytically (\S\ref{sec:design}).

\subsection{The Derivative Accuracy Problem}
\label{sec:derivative-accuracy}

Separability detection relies fundamentally on derivative analysis.
In essence, if $f=g(u)+h(v)$ for two variable groups $u$ and $v$, then the
mixed partials coupling one variable from $u$ and one from $v$ vanish.
Multiplicative separability $f=g(u)\,h(v)$ implies that
$\partial(\log f)/\partial x_i$ is independent of variables outside the
group of $x_i$, and compound-coordinate dependence induces low-rank or
aligned derivative structure.  For example, if $f = g(\prod_k x_k^{a_k})$,
then the weighted gradients $x_i \partial f/\partial x_i$ are collinear with
the exponent vector $(a_1, \ldots, a_n)$, while affine, radial, and centered
coordinates likewise leave characteristic signatures in the gradient field.
All of these tests require either accurate second derivatives (for additive
separability) or accurate gradient-ratio and gradient-alignment information
(for multiplicative, generalized-additive, and compound structure).  As we
demonstrated in Paper~I, on the AI Feynman benchmark NestyNet
improves over Adam-trained~\citep{KingmaBa2015} dense networks (multilayer perceptrons, MLPs) by median factors of approximately $2\,100\times$ ($540\times$
after L-BFGS refinement) for function values, $1\,400\times$ ($450\times$) for
first derivatives, and $780\times$ ($250\times$) for coordinate second
derivatives, with parenthetical values giving the comparison after full-batch
L-BFGS refinement~\citep{Nocedal1980,LiuNocedal1989} initialized from Adam.  For the operator probes most directly
tied to separability and residual tests, the corresponding Adam
(L-BFGS-refined) improvements are approximately $1\,400\times$ ($400\times$) for directional
derivatives, $1\,200\times$ ($370\times$) for mixed partials, $780\times$
($250\times$) for directional curvature, and $870\times$ ($260\times$) for
Laplacians.

When Hessian elements carry $\sim$10\% relative error, testing whether
$\partial^2 f/\partial x_i \partial x_j \approx 0$ becomes unreliable, and the
false positives and negatives noted above corrupt the downstream search.

\begin{figure*}[t]
\centering
\includegraphics[width=\textwidth]{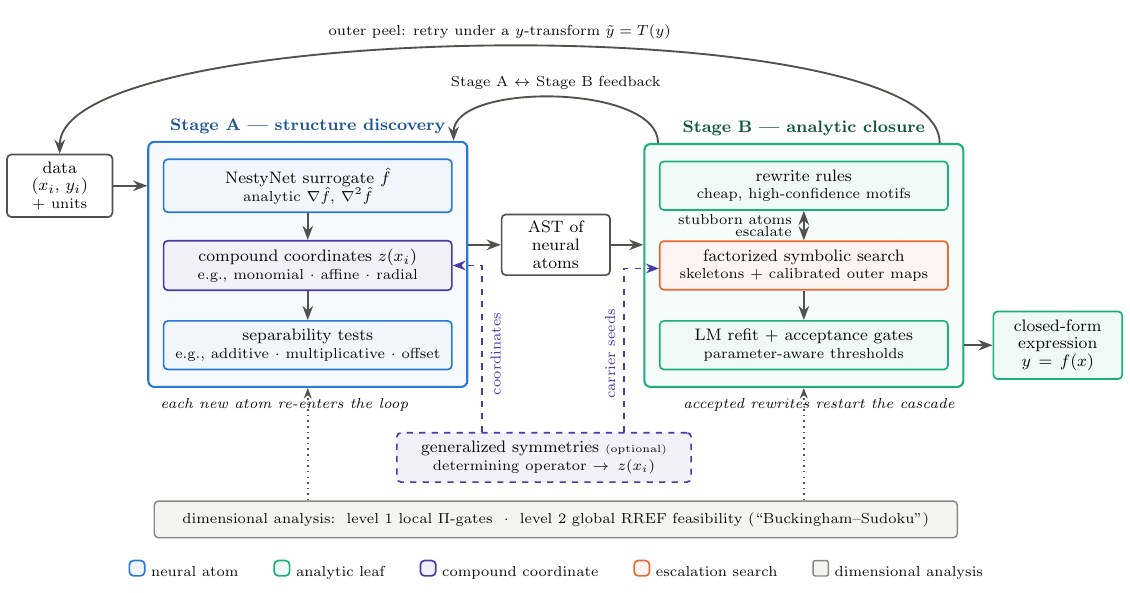}
\caption{Overview of the NestyNet-SR pipeline.  Stage~A trains a NestyNet
surrogate and uses its analytic derivatives to detect separability and
compound coordinates, recursing until every branch of the AST terminates in
a low-arity neural atom.  Stage~B then closes the surviving atoms
analytically, giving first priority to cheap rewrite rules and escalating
stubborn atoms to the factorized symbolic search
(\S\ref{sec:factorized-search}).  The optional generalized-symmetry layer
(\S\ref{sec:gs}, dashed) supplies coordinates to Stage~A and carrier seeds to
the search, dimensional gates (\S\ref{sec:units}) prune candidates in both
stages, and the Y-transforms loop (\S\ref{sec:ytransforms}) retries the
pipeline under a transformed output variable.}
\label{fig:pipeline}
\end{figure*}

\section{Design}
\label{sec:design}

We exploit these analytic derivatives in a two-stage pipeline that
separates concerns.  Figure~\ref{fig:pipeline} gives an overview of the
architecture.  In the first block (Stage~A) we train a NestyNet surrogate $\hat{f}(x)$ to
fit the data, compute mixed partial derivatives analytically, test for
additive, multiplicative, and compound separability, and build an Abstract
Syntax Tree (AST) that records the detected structure.  We then apply this
procedure recursively to each subproblem until every branch of the tree
terminates in a univariate neural atom, or until no further simplification is found.

The second block (Stage~B) then rewrites these neural atoms into closed-form analytical
expressions by orchestrating two complementary mechanisms.  The first is a
library of specialized rewrite rules and template fits for structures such as
separability within composites, homogeneous scaling, ratio invariance, and
other analytic motifs.  These form the cheap, high-confidence first line of
attack.  The second is \emph{factorized symbolic search},
which acts as the general escalation layer for neural atoms that resist
those structured proposals, scoring coefficient-free skeletons by the
calibrated outer maps they admit (\S\ref{sec:factorized-search}).  These two routes share
one acceptance loop, so a stubborn neural atom escalates as soon as its
cheaper candidates are exhausted.  For multi-dataset runs we optionally apply a
post-Stage-B joint refinement on the fixed AST
(\S\ref{sec:class_param_sr}).
The final output is a human-readable analytical expression.

\begin{figure*}[t]
\centering
\includegraphics[width=\textwidth]{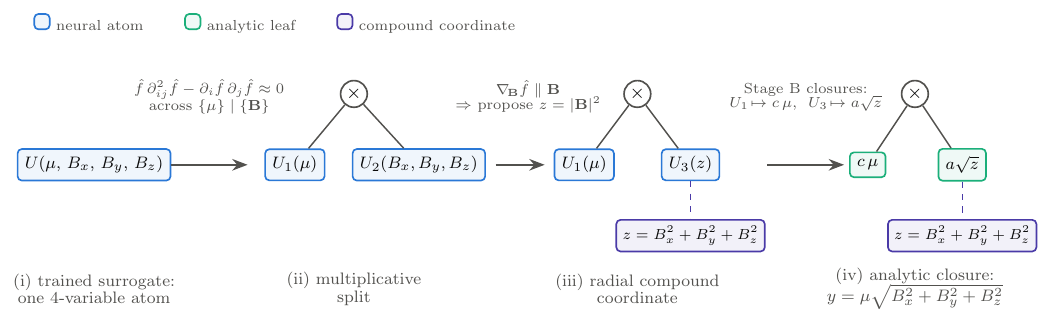}
\caption{The decomposition cascade on AI~Feynman problem \#90,
$y=\mu\sqrt{B_x^2+B_y^2+B_z^2}$.  The multiplicative separability test
(\S\ref{sec:stageA}) splits the four-variable neural atom into
$U_1(\mu)\,U_2(B_x,B_y,B_z)$.  The radial gradient-alignment probe then replaces
the three field components with the compound coordinate
$z=B_x^2+B_y^2+B_z^2$ (its signature is that the
gradient restricted to the field components, $\nabla_{\!\mathbf{B}}\hat f$,
is parallel to $\mathbf{B}$ at every sample, as the chain rule requires for
any $g(|\mathbf{B}|^2)$), and Stage~B closes the remaining univariate atoms as
a monomial and a power law.  Blue atoms are neural, aqua leaves are
analytic, and violet cards are discovered internal coordinates (the same
color roles are used in Figure~\ref{fig:pipeline}).}
\label{fig:filmstrip}
\end{figure*}

Underlying both stages is a single representation.  Every symbolic expression
is a native AST that integrates directly with the
Levenberg--Marquardt (LM) solver~\citep{Levenberg1944,Marquardt1963}.  Structural nodes (binary
Add and Mul, and a Pow node carrying a fixed numeric exponent) and unary
nodes (logs, exponentials, trigonometric and
inverse-trigonometric functions, and a small complex-aware family) compose
with three leaf families, namely input coordinates, fixed constants, and a
single \emph{atom} node.  That atom is a \emph{neural atom} while it wraps a
NestyNet surrogate (its state throughout Stage~A and in the unresolved
branches of Stage~B) and an \emph{analytical leaf} once it has been rewritten
in closed form.  Compound coordinates need no separate node type.  Every atom
carries an explicit tuple of input expressions, each itself an AST of
structural and unary nodes (with no trainable constants inside a
coordinate), so a leaf such as $g(x_0^2/x_1,\,x_2)$ is assembled from ordinary Mul and Pow
nodes, and raw coordinates, monomials, ratios, radial groups, and
trigonometric arguments share one compilation path.  The AST supplies analytic
gradients and Hessians for separability detection and
Jacobian-vector/vector-Jacobian products for the LM solver, and when the tree
is transformed during Stage~B, leaf modules from compatible subtrees are
reused, and incompatible ones warm-start their replacements, so earlier
fitting work is preserved.

This separation of concerns lets each stage be developed
independently, with the AST of neural atoms providing a clean handoff
between surrogate-based structure discovery and symbolic simplification.
A useful consequence is that Stage~A is not tied to our particular Stage~B.
Because it emits a decomposition into low-dimensional neural atoms together
with their discovered internal coordinates, it can serve as a derivative-based
preprocessing front-end for any downstream symbolic regression engine
(genetic programming, reinforcement-learning, or transformer-based), handing
each solver a simpler sub-problem than the original multivariate target.

Following \citet{Tenachi2023PhySO}, NestyNet-SR is able to use physical units
to constrain the search, and it is able to simultaneously fit several
datasets at once~\citep{Tenachi2024ClassSR}, each potentially a different
experiment of the same phenomenon. The algorithm can search for
class-specific parameters (e.g., gravitational acceleration $g$) as well as
experiment specific parameters (e.g., the experiment's spring constant).

This multi-dataset setting also provides a natural bridge from ordinary
symbolic regression to small theory-discovery case studies.  In the vignette
of \S\ref{sec:jacobi_vignette} an ensemble of galactic hosts plays the role
that the ensemble of experiments plays in class symbolic regression.

Our algorithm also supports differential-equation discovery, but we report
that capability separately in the companion paper~\citep[hereafter Paper~IV]{NestyNet2026c} so that the
present article can focus on symbolic regression proper.

\section{Stage A: Separability Detection}
\label{sec:stageA}

Stage~A trains the neural surrogate and uses its analytic derivatives to
decompose the target into structurally simpler pieces, detecting additive,
multiplicative, and compound-variable separability.
Figure~\ref{fig:filmstrip} previews the resulting cascade on a concrete
benchmark equation.  We describe each step in turn.

\subsection{Neural Surrogate Training}

We fit the data with a NestyNet segmented surrogate, whose architecture and
closed-form derivative formulas are given in Paper~I.  Throughout the
present contribution we use a dual-layer composition $f = f^{(1)} \circ
f^{(0)}$, where $f^{(0)}\colon \reals^n \to \reals^M$ maps the inputs to
an intermediate representation and $f^{(1)}\colon \reals^M \to \reals^O$
maps that representation to the outputs, with intermediate dimension set
to $M=n+2$.  This two-layer composition provides substantially greater
expressiveness while preserving closed-form derivative access.

\subsection{Classical Decomposition Motifs}

Following \citet{Udrescu2020,Udrescu2020b}, we test for additive
separability, multiplicative separability, and generalized additivity.
The novelty here is not the motif set itself but the use of analytic
NestyNet derivatives to evaluate it.

For additive separability we test whether the mixed Hessian block vanishes
across a candidate partition, i.e.\ whether
$\partial^2 f/\partial x_i \partial x_j \approx 0$ for $i\in S_1$,
$j\in S_2$.  For multiplicative separability $f=g(x_{S_1})\,h(x_{S_2})$,
differentiating twice gives the equivalent criterion
\begin{equation}
f\,\frac{\partial^2 f}{\partial x_i \partial x_j} - \frac{\partial f}{\partial x_i}\frac{\partial f}{\partial x_j} \approx 0 \, ,
\qquad i\in S_1,\; j\in S_2 \, ,
\end{equation}
which sets to zero the mixed second derivative of $\log|f|$.  The
implementation evaluates that derivative directly as a rational expression
in $f$ and its first two derivatives (no logarithm is taken, so sign
changes in $f$ are harmless) and masks samples where $|f|$ is small.
For generalized additivity in the sense of~\citet{Udrescu2020b}, we test a
derivative criterion equivalent to the gradient-ratio field
$s(x_1,x_2)=\left(\partial f/\partial x_1\right)/\left(\partial f/\partial
  x_2\right)$ factorizing across variable groups, which indicates a wrapped
additive inner structure.  This probe is applied to the surviving
two-variable atoms during the Stage~B rewrites rather than in the initial
Stage~A cascade.

We also test for variable non-dependence, the degenerate case of
separability in which one or more coordinates do not affect the current
leaf.  Operationally, for each neural atom we measure the normalized
gradient magnitude in each effective input direction.  If
$\partial f/\partial x_j$ is negligible across the sampled domain, we propose
a projected neural atom with that coordinate removed, and the proposal is validated by
refitting the full AST.

We extend these classical motifs in several practical ways.
First, non-active coordinates are frozen to dataset medians during the
tests, which suppresses sample-dependent offsets and substantially improves
numerical stability.  Second, we support approximate and overlapping splits
rather than only clean disjoint partitions, so the search can preserve
shared variables that later get resolved by subsequent rewrites.

We also detect \emph{offset-multiplicative} structure of the form
$f(x)=c+g(x_{S_1})h(x_{S_2})$.  The offset $\hat{c}$ is estimated by robust
aggregation across cross-variable derivative pairs and is accepted only when
its scatter is sufficiently small.  This simple extension turns out to be
important in practice for expressions such as $1+x_1x_2$, which are
invisible to a pure multiplicative test.

\subsection{Monomial Compound Coordinates}

Beyond the classical decomposition motifs, Stage~A searches for internal
coordinates $z(x)$ that reduce the remaining dependence to a simpler
univariate or lower-arity problem.  We begin with monomial compounds, which
extend the pairwise product and ratio motifs of \citet{Udrescu2020b} to
general power-law coordinates
\begin{equation}
z = \prod_i x_i^{a_i} \, .
\end{equation}
If $f(x)=g(z)$, then
\begin{equation}
x_i \frac{\partial f}{\partial x_i} = g'(z)\,z\,a_i \, ,
\end{equation}
so the weighted gradients $u_i := x_i \partial f/\partial x_i$ are collinear
with the exponent vector $\vect{a}$.  We therefore form the sampled weighted
gradient matrix $\mat{U}\in\reals^{N\times n}$ and test it for rank~1 via
a singular value decomposition (SVD), accepting when $\sigma_2/\sigma_1$
falls below a threshold.  The dominant
right singular vector yields a continuous estimate of $\vect{a}$, after
which we snap to nearby integer exponent patterns and validate them by
direct fit quality (half-integer powers are recovered downstream, by the
univariate monomial screen and the Stage~B templates).  This \emph{exponent snapping} favors
physically natural coordinates such as $x_1^2/x_2$ over arbitrary real
powers, reflecting the strong prior that exponents in physical laws are
typically small integers or simple fractions.

We also use a log-derivative variant,
$v_i = \partial \log{|f|} / \partial \log{|x_i|} = (x_i/f)\partial f/\partial
x_i$, which is more robust when $f$ spans a wide dynamic range or when an
outer monomial prefactor distorts the raw weighted-gradient test.

\subsection{Generalized Compound-Coordinate Search}

Another Stage-A extension beyond the motif set of \citet{Udrescu2020b} is an
explicit search over richer internal coordinates, since the useful coordinate
is often not a pure monomial but an affine combination, a radius, a centered
coordinate, or a mixed form with shared outer factors.  Each family has a clean
derivative signature.  An \emph{affine} coordinate $z=\sum_i c_i x_i$ (with
$f=g(z)$) makes the sampled gradients rank~1 along $\vect{c}$, generalizing the
binary $x_i\pm x_j$ motifs to multi-axis integer-affine forms.  A \emph{radial}
group has its restricted gradient $\nabla_S f$ parallel to $\vect{x}_S$, the
signature of $g(\sum_{i\in S}x_i^2)$.  A \emph{preferred-origin} probe fits
$\partial_j f\approx\beta_0+\beta_1 x_j$ and seeds the centered coordinate
$z=x_j-x_0$ with $x_0=-\beta_0/\beta_1$.  When a full-variable test fails on a leaf of three or more
inputs, we retry on prioritized subsets (recovering cases where an outer
prefactor shares variables with an internal compound, such as
$f=x_0\,g(x_0x_1/(x_2x_3))$) and probe difference-product coordinates
$(x_i-x_j)x_k$ through the slopes of pairwise gradient ratios.  Competing
proposals are ranked by structural payoff (a visible prefactor, the support
they explain, and the arity they leave behind), with confidence as the final
tie-break, each with a small kind-aware family of wrappers (square or absolute value,
rational stabilization, trigonometric $\sin/\cos$), and a separability-aware
heuristic prefers the variant that most cleanly exposes later decomposition.

These coordinate families are each the invariant of a one-parameter point
symmetry of the target, a view we develop into a
generalized-symmetry layer in \S\ref{sec:gs}.

\section{Factorized Symbolic Search}
\label{sec:factorized-search}

Conventional symbolic regression engines search directly over
complete expressions $f(x)$, so that the inner coordinate, the outer
nonlinearity, and all numerical constants are entangled in one discrete
tree.  The central idea of factorized symbolic-search (FSS) is to factor the 
problem into three separate sub-problems (this factorization is the sense 
in which the search is ``factorized'').  A \emph{skeleton search} proposes 
a constant-free symbolic carrier
\[
    s = s(x_0,\ldots,x_{m-1}) \, ,
\]
an \emph{outer scoring} step fits the best of a small family of nuisance
mappings $M_\theta \in \mathcal{M}$ on $s$, possibly augmented by a
low-complexity additive residual head $h \in \mathcal{H}$, and a final
\emph{continuous refinement} step sharpens the most promising candidates.
The search loss is
\[
    \mathcal{L}(s)
    =
    \min_{M_\theta \in \mathcal{M},\ h \in \mathcal{H}} \|y - M_\theta(s(x)) - h(x)\|^2 + \lambda\, C(s, M_\theta, h) \, ,
\]
so expressions that differ only by a calibrated outer map are scored
through the same underlying carrier rather than being treated as unrelated
trees.  In the configuration used throughout this paper $\lambda$ is zero
and parsimony acts lexicographically instead: among candidates whose errors
agree to within a small factor, the simpler mapping and the smaller tree
are preferred.

\begin{figure*}[t]
\centering
\includegraphics[width=\textwidth]{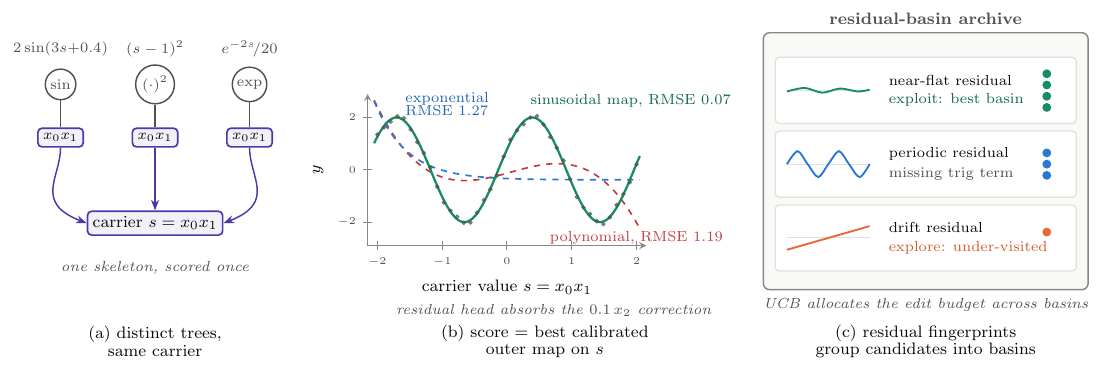}
\caption{The skeleton--mapping factorization of the FSS method, shown on the worked example
$y=2\sin(3x_0x_1+0.4)+0.1\,x_2$ of \S\ref{sec:factorized-search}.
(a)~Expression trees that differ only in their outer rendering share the
carrier $s=x_0x_1$ and are scored through one skeleton.  (b)~The carrier is
scored \emph{after} fitting the best calibrated outer map
(\S\ref{sec:factorized-search:mappings}).  The fitted sinusoidal map alone attains a
root-mean-square error (RMSE) of $0.07$, the floor set by the
not-yet-modeled $0.1\,x_2$ term,
which the additive residual head subsequently absorbs, while the best
cubic-polynomial and exponential maps plateau near RMSE $1.2$.  (c)~Probe
residuals are fingerprinted and grouped into residual basins, and the UCB rule
(\S\ref{sec:factorized-search:mutation}) splits the mutation budget
between top-performing and under-visited basins.  Color roles as in
Figure~\ref{fig:filmstrip}.}
\label{fig:fss_concept}
\end{figure*}

As a concrete illustration, suppose the target is
\[
    y = 2\sin(3 x_0 x_1 + 0.4) + 0.1\, x_2 \, .
\]
A direct tree search must discover the product $x_0 x_1$, the sinusoidal
wrapper, and the frequency, phase, amplitude, offset, and additive
correction in one coupled expression.  In the factorized symbolic-search layer, the
carrier $s = x_0 x_1$ already scores well.  A carrier is never evaluated on its
own.  Its score is the loss $\mathcal{L}(s)$ above, obtained \emph{after} fitting
the best outer map and residual head.  Here that inner fit is trivial (a
sinusoidal outer map fits $A \sin(\omega s + \phi) + c$ and the residual head
absorbs the simple $0.1\, x_2$ correction), so the correct carrier is
rewarded even though the full expression has not yet been assembled.  A
useful consequence is that the fitted nuisance
parameters, here the frequency $\omega$ and phase $\phi$, are later
promoted to numerical constants of the final accepted symbolic form, even
though they were never proposed as discrete tokens during tree search.
More generally, scale, shift, frequency, exponent, and rational-map
parameters can be fitted once a promising coordinate has been exposed, so
that constants that end up deeply nested inside the final expression need
never be discovered by the discrete layer.
Figure~\ref{fig:fss_concept} illustrates the factorization on this example.
Distinct trees sharing the carrier collapse onto one skeleton, the
outer-map battery calibrates the rendering, and residual fingerprints
organize the archive.

This factorization also supplies the steering signal for the search.  A
skeleton receives partial credit whenever some fitted outer map and
residual head explain the target well.  The remaining residual is then
fingerprinted so that archive entries with similar failure modes are
grouped into the same residual basin.  Phase~2 samples from this archive
using a mixture of high-performing and under-visited residual basins, while edit
actions are selected by an upper-confidence-bound (UCB)
rule~\citep{Auer2002} whose reward
is the log-improvement in
effective mean-squared error (MSE), optionally augmented by a bonus for
entering a new residual basin.  Residual
repair proposes additive terms aligned with the current residual, and
inverse repair constructs local pseudo-targets for subtrees of near-miss
expressions.  Thus the search is not driven only by whole-expression
fitness, but by a hierarchy of partial-credit signals that point toward
missing structure before an algebraically correct expression has appeared.

\subsection{Scoring: Outer Maps, Residual Heads, and Residual Basins}
\label{sec:factorized-search:mappings}

For each candidate skeleton $s(x)$, the factorized symbolic-search layer
evaluates it on a fit split and a held-out probe split.  Let
$p=s(x)$ denote the raw skeleton output and let $y$ denote the target
values.  The scorer then fits a small battery of outer mappings.
(i)~A \emph{polynomial} map
$y \approx \sum_{k=0}^{d} c_k z^k$ with $z=(p-\mu_p)/\sigma_p$ is fitted by
ordinary least squares.  (ii)~A \emph{power law} $y \approx a\,p^{b}$ is
fitted by log-linear regression under a consistent-sign assumption.  (iii)~A
\emph{Pad\'e rational} $y \approx P(z)/Q(z)$, where $P,Q$ are low-degree
polynomials with $Q(0)=1$, is fitted by a Sanathanan--Koerner weighted
least-squares loop~\citep{SanathananKoerner1963}.  (iv)~A \emph{sinusoidal}
map $y \approx A\sin(\omega z) + B\cos(\omega z) + c$ has its frequency
$\omega$ selected by coarse-to-fine grid search and its linear coefficients
$(A,B,c)$ solved analytically.  (v)~An \emph{exponential} map
$y \approx a\exp(bz) + c$ has $b$ selected by a coarse-to-fine grid search
and $(a,c)$ by a linear solve.

The scorer is selective.  The cheap families (polynomial, power, and
Pad\'e) act as a prescreen, while the more expensive sinusoidal and
exponential fits are invoked only when hints or competitiveness
thresholds justify the additional cost.  This reduces wasted work on
obviously poor candidates while preserving the ability to recover
periodic and exponential atoms.

The scorer may also attach a small additive linear head
\begin{equation}
h(x) = b_0 + \sum_{j=1}^{K} a_j t_j(x) \, ,
\end{equation}
where the terms $t_j$ are simple unit-consistent raw variables or a few
selected pool terms.  This is primarily a \emph{scoring augmentation}, since it
lets partial skeletons receive credit when the missing structure is a simple
additive correction.

After fitting the outer map (and any residual head), the search engine
fingerprints the probe residual.  Candidates with the same residual
signature are treated as belonging to the same residual basin, and the archive
retains a small elite set per residual basin while preferring simpler representatives
when probe error is effectively tied.  This fingerprinting mechanism is
essential for keeping the search focused on genuinely new structural
directions rather than rediscovering equivalent rewrites through different
mutation paths.

\subsection{Phase~1: Budget-Limited Shallow Enumeration}
\label{sec:factorized-search:enumeration}

Phase~1 performs a brute-force sweep over shallow coefficient-free
expression trees and seeds the archive with easy exact or near-exact
structures.  Trees are generated incrementally, depth by depth.  At depth~1
the tree set consists of the $m$ input variables $\{x_0,\ldots,x_{m-1}\}$.  At
each subsequent depth $d+1$, we apply every unary operator ($\mathrm{neg}$,
$\sin$, $\cos$, $\exp$, $\log$, $\sqrt{\cdot}$, $(\cdot)^2$) to each
depth-$d$ tree, and every binary operator ($+,-,\times,\div$) to each pair
of trees with depth $\le d$.  Algebraically equivalent trees (e.g.\
$x_0 + x_1$ and $x_1 + x_0$) are identified by a canonical string
representation and deduplicated, which keeps the combinatorial explosion in
check.  The enumeration is budget-limited rather than exhaustively
fixed-depth, stopping once
the next depth frontier would exceed the configured expression budget.  Each
candidate is simplified, scored on the fit/probe split, and inserted into
the residual-basin archive.  If a structurally satisfactory result is found
early, the explorer terminates before entering the stochastic phase.

When physical units are available, Phase~1 generates only
dimensionally valid trees.  Trees are bucketed by their output
dimension vector.  Unary operators that require dimensionless arguments
($\sin$, $\cos$, $\exp$, $\log$) draw only from the dimensionless bucket,
addition combines same-dimension trees, and multiplication and division combine
trees whose dimensions sum or subtract to the target.  This dramatically
prunes the search space.  In a typical problem with three input
dimensions, the valid fraction of depth-3 trees may be less than 5\% of all
syntactically valid trees.

Before the general tree enumeration, Phase~1 also runs a battery of
lightweight ``peel'' passes that exploit dimensional constraints to collapse
certain two-layer searches into one-layer searches.  Each pass identifies a
small set of dimensionless monomial carriers~$u$ or variable-ratio pairs,
divides out a known outer factor (Lorentz $1/\sqrt{1-u^2}$, Planck
$1/(\mathrm{e}^{su}-1)$, Gaussian $\mathrm{e}^{-su^2}$, hyperbolic
$\operatorname{sech}$/$\tanh$), and searches for the remaining numerator via
monomial or affine fits.  An inverse-trig peel applies $\sin(y)$ or
$\cos(y)$ to the target and enumerates inner expressions at low depth.
Thanks to dimensional filtering, the carrier set is typically reduced to
2--4 candidates, so the entire presearch battery adds negligible cost but
recovers expressions that are otherwise out of reach of budget-limited tree
enumeration.

\subsection{Phase~2: Archive-Guided Mutation and Repair}
\label{sec:factorized-search:mutation}

Phase~2 extends the search beyond the brute-force proposals by mutating and
repairing promising archive entries.  At each iteration, a parent skeleton
is drawn from the archive using a UCB-guided mixture of exploitation
(sampling from top-performing residual basins) and exploration (biasing toward
less-visited residual basins).  The exploration term is what lets the search
escape the local minima that are the perennial difficulty in symbolic
regression.  By deliberately budgeting attempts on under-visited residual
basins, the method keeps probing structurally distinct hypotheses instead of
polishing a single locally-optimal skeleton.  The core discrete actions are
summarized in Table~\ref{tab:mutation_actions}, and
Algorithm~\ref{alg:factorized-search} summarizes the complete search.

\begin{table}[t]
\caption{Mutation and repair actions available in Phase~2 of the factorized
  symbolic search (\S\ref{sec:factorized-search:mutation}).}
\label{tab:mutation_actions}
\begin{ruledtabular}
\footnotesize
\begin{tabular}{l l}
\pcl{0.22\columnwidth}{Action} & \pc{0.60\columnwidth}{Operation} \\
\hline
\pcl{0.22\columnwidth}{\textsc{Replace}} & \pc{0.60\columnwidth}{Replace a random subtree with a new depth-bounded tree} \\
\pcl{0.22\columnwidth}{\textsc{WrapUnary}} & \pc{0.60\columnwidth}{Apply a unary operator at a random node} \\
\pcl{0.22\columnwidth}{\textsc{AddRand}} & \pc{0.60\columnwidth}{Add a random term, $(s, t) \mapsto s \pm t$} \\
\pcl{0.22\columnwidth}{\textsc{MulRand}} & \pc{0.60\columnwidth}{Multiply by a random factor, $(s, t) \mapsto s \times t$} \\
\pcl{0.22\columnwidth}{\textsc{Prune}} & \pc{0.60\columnwidth}{Remove an internal node, keeping one child} \\
\pcl{0.22\columnwidth}{\textsc{Residual}} & \pc{0.60\columnwidth}{Add the pool term most correlated with the residual} \\
\pcl{0.22\columnwidth}{\textsc{Boost}} & \pc{0.60\columnwidth}{Greedily append the pool terms that most reduce the residual} \\
\pcl{0.22\columnwidth}{\textsc{Crossover}} & \pc{0.60\columnwidth}{Splice a donor subtree from another archive elite} \\
\end{tabular}
\end{ruledtabular}
\end{table}

We emphasize that, despite the ``action'' and ``reward'' vocabulary, Phase~2 is
\emph{not} reinforcement learning.  The actions are selected by the fixed UCB
rule of Eq.~\eqref{eq:ucb}, not by a trained policy network, and no
learned policy weights enter the action selection.  The bandit merely allocates a fixed
edit budget across mutation types based on their observed track record on the
current problem.

The \textsc{Residual} action implements the partial-credit steering signal
introduced at the start of this section.  It scores a pre-evaluated pool of
simple candidate terms against the current residual and proposes the
best-aligned additive repair.

We select among these actions using a UCB multi-armed
bandit,
\begin{equation}
\mathrm{UCB}(a \mid s) = \bar{r}(s,a) + c_{\mathrm{ucb}}\sqrt{\frac{\ln n_s}{n_{s,a}+1}} \, ,
\label{eq:ucb}
\end{equation}
where $\bar{r}(s,a)$ is the mean reward of action $a$ applied to
skeleton~$s$, $n_s$ is the total visit count for $s$, $n_{s,a}$ is the count
for action $a$, and $c_{\mathrm{ucb}}$ is the exploration constant.  The
reward for a mutation that transforms parent MSE $\ell_p$ into child MSE
$\ell_c$ is $r = \log(\ell_p/\ell_c)$, with an optional additive bonus
$+\beta$ for a previously unseen residual basin ($\beta=0$ in the runs reported
here).  An $\varepsilon$-greedy
override (a fixed fraction $\varepsilon=0.10$ of random action choices)
ensures continued exploration.

For difficult near-miss candidates, factorized symbolic search
can switch from global mutation to more targeted local repair routes.
Inverse steering chooses a cut path, approximately inverts the surrounding
context to construct a pseudo-target for that subtree, and ranks local
repairs against that pseudo-target before rescoring the full expression
globally.  This local inversion can be applied recursively to nested
subtrees, each time fixing the surrounding tree and solving the resulting
sub-problem by local symbolic regression.  A separate hole-search mechanism
maintains a frontier of partial repair opportunities and can be mediated by
a route scheduler that decides whether to spend the next budget unit on
build, repair, or abstraction.

Because Phase~2 is stochastic, the method is typically run from multiple
random seeds and the duplicate hits are pooled before Stage~B acceptance.
This multiplicity is inexpensive in practice.  As reported in
\S\ref{sec:aif-benchmark}, nearly half of the problems are solved by the
deterministic presearch in under ten seconds, so the seed replicates matter only
for the minority of hard atoms that reach the mutation search, and the seeds
are trivially parallel.
The search engine monitors stall windows and performs soft restarts while
preserving the archive when progress plateaus.

\subsection{Closure-Native Proposal Layer}
\label{sec:factorized-search:closures}

For atoms involving higher-level operator families (a carrier inside an
envelope, or a numerator--denominator pair), reasoning directly over
expression trees is cumbersome.  We therefore let the search work with
typed structural \emph{roles} (carrier, envelope, numerator,
denominator), enumerating role-to-expression assignments under dimension
and depth constraints.  Each assignment is scored by building the
corresponding design matrix and solving the assignment's outer head
analytically against the true target $y$.  Successful proposals are
admitted into a shared intermediate representation on which a final
linear head is refitted globally and then pruned, so no AST is rendered
prematurely.  The same representation is also used for additive-residual
scoring (\S\ref{sec:factorized-search:mappings}) and for continuous skeleton
refinement (\S\ref{sec:factorized-search:refinement}), so the three mechanisms
communicate through a common object.

\subsection{Continuous Skeleton Refinement}
\label{sec:factorized-search:refinement}

Discrete search proposes \emph{structure}, while the continuous
skeleton-refinement stage sharpens promising structures continuously.  Given
a candidate skeleton, this stage introduces a small number of learnable
scale and frequency parameters at sensitive internal nodes, optimizes them
with L-BFGS, and solves the remaining
linear coefficients analytically at each step.  Solving the linear
coefficients in closed form leaves only a handful of nonlinear parameters to
optimize (a variable-projection reduction), and on this small, dense, and
sometimes non-least-squares reduced objective a quasi-Newton method
(L-BFGS) is a more natural fit than the Levenberg--Marquardt solver used for
the structured full-AST least-squares fits elsewhere, since it needs only gradients
and no assembled Jacobian.  We use multiple random restarts because the
induced objective is still nonconvex.

The refinement is gated.  Only candidates that are already competitive
enter this stage.  The goal is not to replace symbolic
search with continuous optimization, but rather to improve the ranking of
good structural candidates and to recover forms such as shifted/scaled
periodic or exponential atoms that are awkward to discover with purely
discrete edits.  A later linear combination phase can compress additive
basis terms, prune negligible pieces, and return a cleaner symbolic
expression.

When factorized symbolic search is run jointly across multiple
datasets, the continuous refinement stage can share the nonlinear refinement
parameters across datasets while fitting linear coefficients separately per
dataset.  This is particularly useful when the datasets share a structural
law but differ by gains, offsets, or other low-order calibrations.

\begin{algorithm}[t]
\caption{Factorized symbolic search as implemented in Stage~B.}
\label{alg:factorized-search}
\begin{algorithmic}[1]
\REQUIRE Fit/probe atom data, search budgets, mapping family battery $\mathcal{M}$, optional dimension vectors.
\STATE Initialize an archive keyed by residual fingerprints, where each residual basin stores a small elite set.
\STATE \textbf{Phase~1 (shallow enumeration):}
\STATE Generate shallow coefficient-free trees up to the expression budget (with dimensional filtering if units are available).
\FOR{each tree $s$}
  \STATE Evaluate $s$ on fit/probe splits and optionally prescreen.
  \STATE Fit the best outer mapping in $\mathcal{M}$ and, if helpful, a small additive residual head.
  \STATE Fingerprint the probe residual and insert/update the corresponding residual-basin elites.
  \IF{a structurally satisfactory candidate is found}
    \RETURN top-$k$ archive rows.
  \ENDIF
\ENDFOR
\STATE \textbf{Phase~2 (archive-guided search):}
\FOR{$t = 1, \ldots, n_{\mathrm{iter}}$}
  \STATE Select a parent residual basin using a mixture of low-loss elites and under-visited residual basins.
  \STATE Form the action slate from admissible build, residual, crossover, closure, inverse-repair, and hole-repair routes.
  \STATE Choose an action by UCB using log-MSE improvement and residual-basin novelty as reward, then apply residual, inverse, dimensional, and prescreen gates.
  \STATE Propose a child or local repair, rejecting it if depth or dimensional constraints fail.
  \STATE For closure proposals, build a basis-state candidate, globally refit/prune its head on $y$, then score it with the same mapping+head pipeline.
  \STATE Harvest unit-compatible subexpressions from competitive candidates into an auxiliary atom pool, retain a small registry as future seed blocks, and directly score local linear spans over observed atoms.
  \STATE Optionally run continuous skeleton refinement on competitive candidates and admit any improved refined/basis-transition variants.
  \STATE Update route statistics and perform soft restart if the search stalls.
\ENDFOR
\RETURN Top-$k$ archive rows, each carrying its skeleton, fitted mapping metadata, and AST embedding.
\end{algorithmic}
\end{algorithm}

\section{Generalized Symmetries}
\label{sec:gs}

The coordinate families of \S\ref{sec:stageA} are each the invariant of a
one-parameter point symmetry of the target.  Monomials arise from scalings,
affine coordinates from translations, radii from rotations, and their
centered forms from affine conjugates.  We have therefore implemented a
generalized-symmetry (GS) layer in NestyNet-SR that discovers the symmetries
themselves, and with them the reduced coordinates.  The layer recovers the
Stage~A detector families (monomial, affine, radial, indefinite quadratic,
difference-product, and centered coordinates) from one operator, replacing a
catalog of special cases by geometry.  More usefully, it reaches coordinates
that lie outside the integer vocabulary of the baseline detectors, and it
hands those coordinates to the factorized symbolic search of
\S\ref{sec:factorized-search} as carrier proposals, closing problems whose
internal coordinate the structural search cannot assemble on its own.

\subsection{The Determining Operator}
\label{sec:gs_determining}

Consider the graph $y=f(x)$ of the surrogate, with $x\in\reals^n$, and the
affine vector field
\begin{equation}
V \;=\; (\mat{A}x+\vect{b})\cdot\nabla_x \;+\; (\alpha+\beta y)\,\partial_y \, ,
\label{eq:gs_generator}
\end{equation}
with an unknown input action $(\mat{A},\vect{b})$ and an unknown output
action $(\alpha,\beta)$.  The field $V$ generates a one-parameter family of
transformations that maps the graph of $f$ to itself precisely when $V$ is
tangent to the graph, which is the determining equation of Lie
symmetry analysis~\citep{Olver1993},
\begin{equation}
\nabla f(x)\cdot(\mat{A}x+\vect{b}) \;-\; \alpha \;-\; \beta f(x) \;=\; 0 \, .
\label{eq:gs_determining}
\end{equation}
The essential observation is that Eq.~\eqref{eq:gs_determining} is linear in
all $n^2{+}n{+}2$ unknowns jointly, so every sampled point contributes one
row of a design matrix, and the full Lie algebra of affine symmetries of the
target is the nullspace of the stacked system, obtained by a single SVD.
The analytic NestyNet gradients make these rows accurate enough for the
construction to be practical.

Two safeguards are imposed to ensure the recovered algebra is trustworthy.  
The scientific object is the nullspace \emph{subspace}, not any
individual basis vector (when the nullity exceeds one, an SVD basis is
arbitrary), so all downstream logic operates on the subspace projector, the
rank of the pointwise distribution it spans, held-out residuals, and
bootstrap principal angles.  A candidate algebra must further pass
certificates (closure of the Lie bracket including the output action, a
stable distribution rank, a consistent quotient dimension) before any
coordinate is built from it.  A certified algebra is then compiled into a
reduction plan: invariant coordinates $z(x)$ spanning the quotient, an orbit
coordinate $s$, and an output normal form.  For an accepted generator with
$Vf=\alpha+\beta f$, the target takes the form
\begin{equation}
\begin{gathered}
f = -\frac{\alpha}{\beta} + e^{\beta s}\,H(z) \quad (\beta\neq 0) \, ,\\
f = \alpha s + H(z) \quad (\beta=0) \, ,
\end{gathered}
\label{eq:gs_normalform}
\end{equation}
so a promoted reduction proposes the inner coordinates and the outer
structure simultaneously.  Promotion itself is gated, so that a reduction enters the
active Stage~A slate only if it survives a complexity-penalized comparison
against the unreduced expression at matched held-out accuracy, and under the default
policy it only augments the baseline proposals.

\subsection{Charts, Composition, and Discovered Warps}
\label{sec:gs_charts}

The determining operator is affine, but a fixed change of variables extends
its reach.  Solving Eq.~\eqref{eq:gs_determining} in the chart
$u_i=\log x_i$ (with chain-ruled gradients) turns scalings into
translations, so monomial coordinates $\prod_i x_i^{a_i}$ appear as linear
invariant covectors whose components are snapped to small rationals and
revalidated against the determining residual.  The chart $u_i=1/x_i$
similarly exposes coordinates linear in reciprocals, $\sum_i c_i/x_i$.  Rather
than enumerating charts, a further step \emph{discovers} the per-axis warp
that linearizes a hidden generalized-additive symmetry
$f=g\bigl(\sum_i c_i\,\varphi_i(x_i)\bigr)$.  The certificate is the
normalized off-diagonal Hessian ratio
$R_{ij}=(\partial_i\partial_j f)/(\partial_i f\,\partial_j f)$, which equals
$g''/g'^2$ for every pair $(i,j)$ whenever such a warp
exists, a necessary signature that the reconstruction below
then confirms.  The
warps $\varphi_i$ are then recovered from pairwise log-gradient differences
(the whole power family snaps in a single linear fit) and the covector
$c_i$ falls out of a rank-one test.  This reaches mixed-power coordinates
such as $x_0^2+x_1^3+x_2^2$ that no fixed chart exposes.

On trained surrogates, whose gradients carry relative errors of order
$10^{-3}$, the global nullspace solve can lose bracket closure.  The layer
therefore also composes coordinates from \emph{pairwise witnesses},
one-dimensional scaling, translation, rotation, and Lorentz-boost tests
that remain reliable at surrogate noise levels, chaining accepted pairs
into global monomial, linear, radial, and indefinite-quadratic coordinates
such as $x_0^2-x_1^2-x_2^2-x_3^2$.  A promoted coordinate can serve as a
virtual axis, so nested forms such as $(x_0x_1/x_2)-x_3$ compose as well,
and a noise-calibrated tier of acceptance statistics lets certified
reductions be promoted from trained surrogates.

\subsection{Coordinate Discovery Beyond the Detector Vocabulary}
\label{sec:gs_oblique}

\begin{figure*}[t]
\centering
\includegraphics[width=\textwidth]{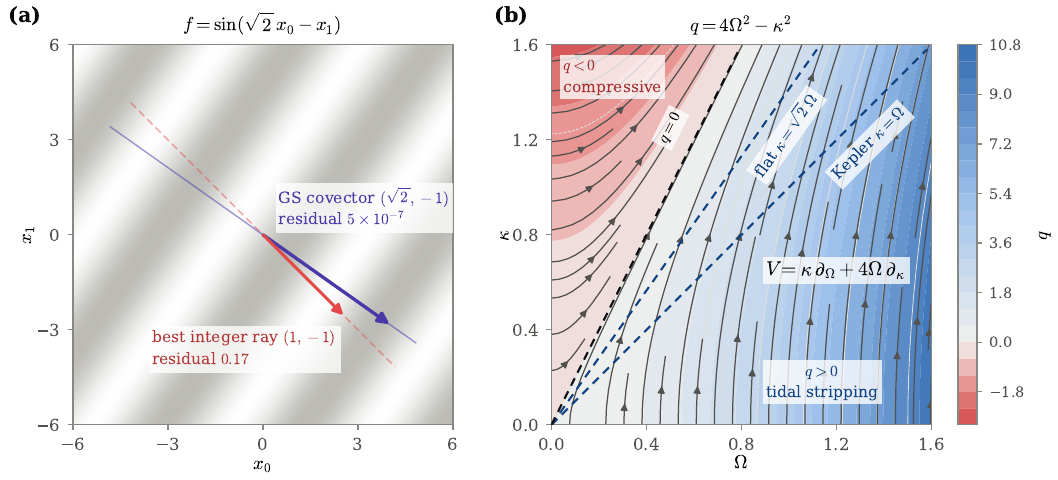}
\caption{The geometry seen by the generalized-symmetry layer.
(a)~For $f=\sin(\sqrt{2}\,x_0-x_1)$ (gray stripes), the best ray available
to the integer-affine detector, $(1,-1)$ (red, dashed), is misaligned with
the stripe normal by $\Delta\theta\simeq0.17$~rad, while the GS affine
solve recovers the oblique covector $(\sqrt{2},-1)$ (violet) with
determining residual $5\times10^{-7}$ (\S\ref{sec:gs_oblique}).
(b)~The $(\Omega,\kappa)$ plane of the tidal-radius example
(\S\ref{sec:jacobi_vignette}).  The flow of the discovered generator
$V=\kappa\,\partial_\Omega+4\Omega\,\partial_\kappa$ (streamlines) is
everywhere tangent to the level sets of its invariant
$q=4\Omega^{2}-\kappa^{2}$ (blue/red fill), the defining property
$Vq=0$ made visible.  Dashed rays mark the Keplerian and
flat-rotation-curve limits and the $q=0$ cone separating stripping from
compressive tidal regimes.}
\label{fig:gs_geometry}
\end{figure*}

The baseline Stage~A detectors search small integer (or half-integer)
coefficient patterns.  The GS layer solves for arbitrary real generators, so
it can recover coordinates the baseline cannot express.  For the target
$\sin(\sqrt{2}\,x_0-x_1)$ of Figure~\ref{fig:gs_geometry}a, the
integer-affine detector proposes its best
integer ray with a coordinate-angle residual of $0.17$, while the GS affine
solve recovers the oblique covector to $5\times10^{-7}$.  On a generic
target with no symmetry, the solver correctly proposes no generator at any
tested noise level.  We distinguish two levels of success.  Recovering a
known detector family counts only as \emph{detector recovery}, whereas the
oblique case is a genuine \emph{coordinate discovery}, since the coordinate was
not in the baseline vocabulary at all.
Figure~\ref{fig:gs_geometry}b previews the generator flow of the
anisotropic tidal invariant discovered in the tidal-radius example of
\S\ref{sec:jacobi_vignette}.

\subsection{Supplying Carriers to the Factorized Search}
\label{sec:gs_fss}

The factorized search of \S\ref{sec:factorized-search} separates
structure from calibration, but it must still assemble the internal
coordinate by discrete search, and deeply structured carriers (signed
quadratics, four-variable monomials, mixed-power sums) are exactly where it
fails (\S\ref{sec:aif-benchmark}).  The GS layer bypasses that assembly.
It discovers the coordinate from the gradient geometry of the
target and seeds it into the search as a carrier, after which the ordinary
outer-map battery of \S\ref{sec:factorized-search:mappings} fits $g(z)$
in closed form.  Table~\ref{tab:gs_fss} shows illustrative carrier-consumption
cases at a matched budget in oracle mode (exact function values, no surrogate,
\S\ref{sec:aif-benchmark}).  In every non-control case FSS consumes the
discovered carrier and closes the problem at the seeding step.  The fifth row
is AI~Feynman problem \#3, one of the targets of the oracle benchmark of
\S\ref{sec:aif-benchmark}.  On the non-symmetric control the GS layer
emits no seed and FSS continues through its ordinary search path.  

The seed closes problems of the form $y=g(z(x))$.  When a prefactor 
multiplies the coordinate, as in $x_0\sqrt{x_1^2+x_2^2+x_3^2}$ 
(AI~Feynman problem \#90), the coordinate alone does not suffice.  
The full pipeline handles that case through the Stage~A reduction machinery 
(prefactor peeling and atom arity reduction), which is how problem 
\#90 is solved in the full-pipeline benchmark of \S\ref{sec:full}.

\begin{table}[t]
\centering
\caption{Illustrative GS carriers consumed by factorized symbolic search
(oracle mode, $2\,500$ mutation iterations, single seed).  Values are
held-out probe MSE.  In each non-control case FSS scores the GS coordinate
at the seeding step and the outer-map battery closes the fit.  On the
control target GS emits no carrier and FSS continues through its ordinary
search path.}
\label{tab:gs_fss}
\begin{ruledtabular}
\footnotesize
\setlength{\tabcolsep}{2pt}
\begin{tabular}{lc}
Target & FSS \\
\hline
$\sin(x_0^2-x_1^2-x_2^2-x_3^2)$ & $5.1\times10^{-13}$ \\
$\sin(x_0x_1x_2/x_3)$ & $1.8\times10^{-12}$ \\
$\sin((x_0-x_1)^2+(x_2-x_3)^2)$ & $1.1\times10^{-13}$ \\
$\sin(x_0^2+x_1^3+x_2^2)$ & $6.6\times10^{-13}$ \\
$\sqrt{(x_0-x_1)^2+(x_2-x_3)^2}$ & $1.2\times10^{-30}$ \\
$\sin(x_0)\,x_1+x_2\cos(x_3)$ (control) & no GS carrier \\
\end{tabular}
\end{ruledtabular}
\end{table}

\section{Stage B: Hybrid Rewrites and Search Escalation}
\label{sec:stageB}

Stage~B takes over once Stage~A has reduced the problem to neural atoms.  It
is itself also a tiered symbolic-search system, giving first priority to
specialized rewrite rules for structural patterns that admit cheaper and more
reliable targeted proposals than open-ended combinatorial search, and
escalating to the factorized symbolic-search engine of
\S\ref{sec:factorized-search} when those structured routes do not close the
problem cleanly.
(The Stage~A separability tests are reused inside Stage~B to expose
newly-uncovered structure, but the two stages remain distinct blocks.)

\subsection{Two-Phase Iterative Engine}
\label{sec:stageB_engine}

The Stage~B engine runs in two phases that enact an effective Occam's razor
preferring simple separable structure.  Phase~1, the ``separability loop'',
exhaustively applies structure-discovering rules until convergence, peeling
away every additive, multiplicative, subtree, counterterm, counterfactor, and
log-ratio split it can find, together with the bounded univariate closures
(which must be found before a generic approximant can consume a leaf), before
handing the residual neural atoms on.
Phase~2 then applies the specialized analytical rewrites (multivariate
compression, homogeneity peeling, ratio invariance,
compound-function macros, outer transforms) to whatever atoms remain neural.
Whenever it accepts a rewrite the engine restarts from Phase~1, so
newly-exposed structure is fully decomposed before more complex forms are
tried.  (These two Stage~B phases are unrelated to the two phases of the
factorized search of \S\ref{sec:factorized-search}.)

\subsection{Rewrite Rules}
\label{sec:stageB_rules}

In essence, Stage~B is implemented as a rule-driven search over AST
rewrites.  Each rule proposes a finite set of candidate replacements for one
(sub)expression, which are then validated by refitting and a parameter-aware
acceptance test (\S\ref{sec:stageB_accept}).  The suite includes rules for 
neural-leaf and subtree separability, homogeneity and
product-homogeneity peeling, ratio invariance and coupled-ratio factoring,
compound-function macros, additive polynomial splitting, univariate and
multivariate neural-network compression, log-ratio detection, counterterm and
counterfactor splits, and outer-transform and preconditioner fallbacks.

\subsection{Compound-Function Macros}
\label{sec:cfmacros}

Many physical expressions involve recognizable analytic motifs that resist
decomposition into simple polynomial, trigonometric, or rational forms
($\mathrm{sinc}$-like factors, $\sin(az)/\sin(z)$ ratios, square-root branch
factors $\sqrt{u^2+v^2}$), and which are brittle and combinatorially expensive
to assemble from primitives.  We therefore keep a small library of
\emph{compound-function macros}, high-payoff templates that expand to ordinary
AST expressions but are proposed as atomic rewrite moves, screened by
closed-form least squares and confirmed by full LM fits.  This plays a role
analogous to the fixed special functions of AI~Feynman~2.0~\citep{Udrescu2020b},
except that our motifs are screened template fits rather than brute-force
matches.  For shared-prefactor templates $y=p(x)\,[\,b+a\,m(x)\,]+c$ the
screening uses the feature matrix $(p\,m,\,p,\,1)$ in $y$-space, which is a 
stable three-parameter regression that identifies forms such as
$y=p(x)[\sqrt{1+u(x)}\cos\Delta(x)+1]$ or $y=p(x)\,\mathrm{sinc}^2(\phi(x))$.

\subsection{Candidate Acceptance Logic}
\label{sec:stageB_accept}

Each proposed rewrite becomes a \emph{Candidate} (new AST, initialization,
metadata) that is first screened by static prechecks, rejecting duplicates,
non-tree structures, and dimensionally inconsistent forms before any fitting.
Survivors are fitted by LM and accepted by a layered rule.  With the default
heuristic acceptance logic (distinct from the statistical selection of \S\ref{sec:stat_selection}), a candidate is
accepted unconditionally when the loss improves beyond a tolerance, and
separability rewrites are granted a small additional loss budget (of order
ten percent), since structure discovery is worth a small accuracy cost.
Cosmetic simplifications that cut
the parameter count $P_{\rm cand}$ below $P_{\rm base}$ are accepted under a
parameter-aware ceiling scaled by
$10^{0.30\,\max(0,\log_{10}(P_{\rm base}/P_{\rm cand}))}$, the whole ceiling
being clamped to an absolute cap tied to the acceptable-loss scale, and once
both losses sit below the noise floor the comparison switches from loss
ratios to simplicity, so that near-exact rewrites are not rejected on
noise.

\subsection{Statistical candidate selection}
\label{sec:stat_selection}

By default the code selects competing candidates for simplification based on
the measured loss with heuristic bonuses for simpler expressions. An optional
stricter method is also provided, which replaces the heuristic 
score by a confidence-aware complexity--accuracy frontier.

This statistical mode separates discovery from selection. At the outset, an
audit sample is sealed and hidden from all proposal and polishing steps. Once
the search phase ends, duplicate analytic expressions are merged and the 
archive is frozen.  Every candidate is then scored on each declared
independent audit unit using the same bounded loss, with fitted coefficients 
held fixed and invalid evaluations receiving a fixed penalty.

A complexity vector records expression size, depth, fitted coefficients, 
and numerical-constant description length.  For candidates \(i\) and \(j\), we 
form paired loss differences \(d_{ij,u}=\ell_{i,u}-\ell_{j,u}\).  Candidate \(i\) 
removes \(j\) from the strict confidence Pareto front only when \(i\) is no more 
complex in every component of the complexity vector and the upper
bound on \(\mathbb{E}[d_{ij}]\) is negative.  We permit a
strictly simpler expression to dominate when this bound does not exceed a
tolerance \(\delta\).  The bounds are one-sided simultaneous
$1-\alpha$ bounds ($\alpha=0.05$) over the pre-declared set of
admissible ordered pairs, obtained from a Gaussian-multiplier max-$t$
bootstrap of the paired unit-level differences, with a
one-sided Bonferroni $t_{1-\alpha/K}$ critical value substituted whenever the
unit count or family size lies outside the bootstrap's calibrated range.

A separate identification step reports the final law. Beginning with the simplest
candidate in a pre-declared total order (free parameters, then
constant description length, node count, and depth), it accepts extra
complexity only when the lower confidence bound on improvement,
drawn from the same simultaneous family, exceeds \(\delta\)
(the runs reported here adopt $\delta=0$).  Thus the audit is consulted only after the
hypotheses are fixed, protecting against selection overfitting.  We note that 
the result is conditional on the frozen archive and searched grammar.

\subsection{Committee-of-experts validation mode}
\label{sec:coe}

Although the algorithm can make decisions based on a single split of training 
and validation data, we provide the option to employ a small deterministic 
committee instead.  One reference slice proposes the Stage-A and 
Stage-B modifications, while the remaining slices act as witnesses that
re-evaluate each incumbent--candidate comparison on data the reference never
saw (both incumbent and candidate are refit on the witness slice).  Each
witness votes candidate, incumbent, or tie according to whether
$\ell_{\rm cand}-\ell_{\rm inc}$ falls below, above, or within a noise-aware
tolerance
\begin{equation}
\tau =
\max\!\left[
\eta\,\sigma_{\rm noise}^2\sqrt{\tfrac{2}{N_{\rm val}}},\;
\epsilon_{\rm rel}
\max(\ell_{\rm inc},\ell_{\rm cand},\sigma_{\rm noise}^2)
\right],
\end{equation}
with $\eta,\epsilon_{\rm rel}$ user constants.  Ties never veto, and in the
final committee adjudication they are broken toward the simpler expression.
When run in combination with the statistical selection of \S\ref{sec:stat_selection},
the committee only serves to provide additional proposals and diagnostics.
The campaigns of \S\ref{sec:full} use eight scout proposers, run on
separate data slices, whose candidates join a reservoir that 11 witness 
slices then evaluate.

%

\subsection{Gauge Fixing and Gauge-Aware Rewrites}
\label{sec:stageB_gauge}

Symbolic decompositions are often non-identifiable, in that many parameter
settings, or even many AST representatives, denote the same function.  Left unmanaged
this proliferates equivalent candidates and destabilizes the fits, so we treat
gauge fixing as part of the search space.  \emph{Scalar} gauges are removed by
compiling fitted leaves into reduced parameterizations.  A product of trainable
factors carries a single shared dimensionless scale with each factor's leading
coefficient pinned to $+1$.  Exponential-polynomial, rational, and
log-polynomial leaves have the analogous multiplicative or quotient gauge,
resolved by pinning a constant term or normalizing the denominator on a stable
pivot.  Additive sibling leaves with canceling medians are recentered to a
balanced representative, which also protects the Hessian-based separability
tests of \S\ref{sec:stageA} from large canceling offsets.

Overlapping decompositions, where two leaves share variables, carry a stronger
functional gauge.  In $F(s,u,v)=g(s,u)+h(s,v)$ any $\phi(s)$ transfers
between the leaves (multiplicatively, $g\mapsto g\,\psi(s)$,
$h\mapsto h/\psi(s)$).  We anchor these at reference slices (dataset medians),
constraining one leaf to vanish or be constant there via a soft penalty ramped
after an initial unconstrained fit.  Crucially, an overlapping split is
pre-tested \emph{gauge-invariantly}, before any fitting, by the four-corner
identity
\[
    F(s,u,v)-F(s,u,v_0)-F(s,u_0,v)+F(s,u_0,v_0) \approx 0
\]
(with the ratio analogue
\[
F-F(s,u,v_0)\,F(s,u_0,v)/F(s,u_0,v_0) \approx 0
\]
for multiplicative overlaps), the
residual normalized by the leaf's output scale and thresholded against the
surrogate's own probe residual, so unsupported splits are rejected cheaply.
The rule scheduler is likewise gauge-aware, giving scope-resolving transfers
and counterterm/counterfactor splits priority over leaf-local compression so
the search does not lock in a neat representative of the wrong global gauge.
Together these roles, better conditioning, removal of non-identifiable degrees
of freedom, and protection against gauge-equivalent rewrites, keep the search
converging on the simplest global structure.

\subsection{Univariate NN Compression}

For univariate atoms $\texttt{U}(x_i)$ we fit a family of analytical
templates spanning monomials and low-degree polynomials, exponential and
Planck- and sech-like physics forms, logarithmic and shifted-log forms,
algebraic and rational forms (roots, Pad\'e, exp-polynomial, exp-rational),
bounded $\tanh$ nonlinearities, and periodic (trigonometric) templates,
seeded where possible by the Stage-A feature priors of
\S\ref{sec:features}.  The scaling-feature seeds are not limited to pure
monomials.  Depending on the inferred homogeneity exponent they propose
low-complexity linear, quadratic, half-power, or reciprocal-rational leaves.
Each candidate is fitted by LM optimization and evaluated by the acceptance
logic of \S\ref{sec:stageB_accept}.

\subsection{Trapped Variable Factorization}

For multivariate NN atoms where standard separability fails, we attempt
\emph{trapped factorization}, that is, finding a multiplicative
decomposition where one factor depends on a subset of variables,
\begin{equation}
f(x_1, x_2, \ldots) = g(x_1) \cdot h(x_1, x_2, \ldots) \, .
\end{equation}
Detection uses polynomial fitting along one-dimensional slices and checking
for consistent factorization structure across the domain.  It transpires
that this ``trapped'' pattern arises more often than one might expect,
because many physical laws contain an envelope factor that depends on fewer
variables than the full expression.  Alongside trapped factorization, the
multivariate compression rule fits quadratic and higher-degree polynomials,
rational and square-root (rational) forms, log- and exp-rational forms,
generalized-additive compositions $\phi(\mathrm{NN}_1+\mathrm{NN}_2)$, affine
splits, and trigonometric envelopes and difference forms.

\subsection{Post-Stage-B Multi-Dataset Parameter Refinement}
\label{sec:class_param_sr}

When multiple datasets share the same discovered structure but not
necessarily the same constants, we run an optional post-pass after Stage~B.
The AST structure is kept fixed, and each leaf tag is assigned either
\emph{class} scope (shared across datasets) or \emph{experiment} scope
(dataset-local).  

By default, the class/experiment split is initialized from per-tag parameter
dispersion in the Stage~B fits (a coefficient-of-variation threshold), then
refined by greedy one-tag-at-a-time acceptance against aggregated validation
loss.

The joint fit minimizes the residual error,
\begin{equation}
\mathcal{L}_{\mathrm{fit}}
=
\sum_{d=1}^{D} w_d\,
\left\|
\phi\!\left(y^{(d)}\right)-\phi\!\left(\hat{y}^{(d)}\right)
\right\|_2^2,\qquad
w_d=\frac{N_d}{\sum_j N_j},
\end{equation}
where $\phi$ is the same optional data transform used by Stage~B (identity, asinh,
etc.), and $N_d$ is the number of points in dataset~$d$.

To address reparameterization and gauge effects (situations where physically
shared quantities are combinations of leaf parameters rather than single
leaves) we add a derived-invariant discovery step\footnote{Named
  ``Parameter-SR'' in the codebase.}.  It scans scalar references from
trainable leaf parameters and fixed scalar buffers and evaluates candidate
derived invariants of the forms $s_a s_b$, $s_a/s_b$, and $s_a s_b^2/s_c$
across datasets.  Low-scatter candidates are retained using a robust
median-absolute-deviation (MAD) score and can be enforced as soft
constraints,
\begin{equation}
\mathcal{L}
=
\mathcal{L}_{\mathrm{fit}}
 +
\lambda\,
\frac{1}{K}
\sum_{k=1}^{K}
\frac{1}{D}
\sum_{d=1}^{D}
\left(
\frac{I_k^{(d)}-\bar{I}_k}
{\operatorname{median}_j\!\left|I_k^{(j)}\right|+\epsilon}
\right)^2.
\end{equation}
Here $I_k^{(d)}$ is derived invariant $k$ evaluated on dataset $d$, and
$\bar{I}_k$ is its dataset mean.  We found that even a modest number of
discovered invariants can substantially stabilize the joint fit,
particularly when the number of datasets is small relative to the number of
free parameters.

\section{Feature Discovery}
\label{sec:features}

The rule proposals in Stage~B are guided by a battery of cheap structural
probes run on the trained surrogate.  All share one logic, sampling the
surrogate and its analytic gradients and emitting a hint only when the relevant
signature has low scatter across the domain.  The exemplar is
\emph{homogeneity} detection over a variable subset $S$, from the restricted
Euler relation
\begin{equation}
\sum_{i \in S} x_i \frac{\partial f}{\partial x_i} \approx k_S \, f \, ,
\end{equation}
whose pointwise ratio $r_S(x)=\big(\sum_{i\in S}x_i\,\partial_i f\big)/f(x)$ is
accepted as a homogeneity degree $k_S$ when its relative scatter is small.
Because the test runs over all subsets, it exposes partial homogeneity such
as $f=x_1^2\,g(x_2)$.  A direct rescaling probe complements it, fitting the
slope of $\log(f(\lambda z)/f(z))$ against $\log\lambda$ per axis and along
any compound coordinate $z$ already introduced by Stage~A (including the
simple forms $x_i\pm x_j$, $x_ix_j$, $x_i/x_j$).

The remaining probes follow the same accept-on-low-scatter pattern.
Trigonometric detection grids over candidate frequencies, tests
$\sin(\omega z)$, $\cos(\omega z)$, and $1-\cos(\omega z)$ carriers, and
resolves product- versus
difference-argument structure by regressing the per-slice frequency and
phase against the partner coordinate.  Radial and preferred-origin detection reuse the
gradient-alignment and shift-fitting probes of \S\ref{sec:stageA}, the latter
seeding centered-coordinate proposals.  Parity is probed independently by
reflection tests about each axis median.  Saturation detection flags monotonic axes that flatten in the tails,
suggesting rational or $\tanh$ rewrites.  The scaling, trigonometric, and
saturation hints gate which Stage~B rules fire, while the remaining probes
are recorded as diagnostics.

\section{Y-Transforms}
\label{sec:ytransforms}

When separability detection fails in the original output space,
transforming the dependent variable can reveal hidden structure.
This situation arises frequently in physics.  For example,
$y = \exp(g(x))$ is not additively separable in $y$ but becomes so after
a logarithmic transform.

We maintain a registry of 15 output transformations organized into five
families: (i)~identity ($\tilde{y} = y$), (ii)~power transforms
($\sqrt{y}$, $y^2$, $1/y$, and $\tilde{y}=y^2-1$, i.e.\
$y=\sqrt{1+\tilde{y}}$),
(iii)~logarithmic transforms ($\log(y)$ and $\log(-y)$ for negative
data), (iv)~exponential transforms ($\exp(y)$ and $\exp(-y)$), and
(v)~trigonometric and inverse-trigonometric transforms ($\sin(y)$,
$\cos(y)$, $\tan(y)$, $\arcsin(y)$, $\arccos(y)$, $\arctan(y)$).  Each
transform specifies forward and inverse operations, first and second
derivatives for chain-rule propagation through separability tests,
domain constraints (e.g., $\log$ requires $y > 0$), and units
compatibility (e.g., $\log$ is incompatible with dimensioned
quantities).  The separability tolerances remain meaningful across transforms because
the tests normalize by the scatter of the transformed output, so a
transform that amplifies small variations (such as $\log y$) is judged on
its own scale.
Rather than retraining under every transform, we screen all candidates
through a chain-rule virtual model that exposes $T(f(x))$ and its
derivatives from the already-trained identity surrogate, rank the
transforms by the structure these screens reveal, and rerun Stage~A only
on the leading few.  If separability is detected for some $T$, we discover
the expression $\tilde{f}(x)$ and report $y = T^{-1}(\tilde{f}(x))$.

\section{Dimensional Analysis}
\label{sec:units}

NestyNet-SR optionally enforces dimensional consistency at two levels.
\emph{Local} checks validate individual operations, and a \emph{global}
constraint propagation system detects dead-end configurations before
expensive fitting. We describe each level in turn, beginning with the units
specification that both levels share.

\subsection{Units Specification}
\label{sec:units:spec}

Each quantity is represented by a \emph{dimension vector}
$\vect{u}\in\mathbb{Q}^B$, where $B$ is the number of base dimensions (e.g.\
$B=3$ for a $[L,T,M]$ basis),
\begin{equation}
[y] = [L^{a_1} T^{a_2} M^{a_3}] \;\leftrightarrow\; \vect{u}_y = (a_1, a_2, a_3) \, .
\end{equation}
The user supplies $\vect{u}_y$ for the target, $\vect{u}_{x_i}$ for each
input variable, and optionally $\vect{u}_c$ for free or fixed constants. All
exponents are exact rationals, so that fractional powers like
$\sqrt{\cdot}$ are handled without floating-point error.

\subsection{Local Consistency Checks (Level~1)}
\label{sec:units:local}

Four local gates validate individual operations during the search. For
addition, both operands must have matching dimensions. For multiplication,
dimension exponents add, $[\text{parent}] = [\text{left}] +
[\text{right}]$. For exponentiation,
$[\text{parent}] = s \cdot [\text{base}]$ where $s\in\mathbb{Q}$ is the
exponent. For transcendental functions ($\log$, $\exp$, $\sin$, $\cos$), the
argument must be dimensionless and the output is dimensionless.

These checks are applied at each structural decision point: full-AST
validation after Stage~B rewrites, Buckingham $\pi$-count verification for
compound-variable proposals, single-node feasibility checks for
additive/multiplicative splits, and output-dimension inference for
individual atoms. Each gate examines its own local neighborhood (``Is this
compound/split/rewrite dimensionally legal in isolation?'') but it does not
ask whether, given this choice, the \emph{rest} of the tree can still be
satisfied.

\paragraph{Compound Dimensional Rank Gate.}
\label{sec:units:local:compound-rank}

When a compound variable $z = \prod_i x_i^{a_i}$ replaces a subset of an
atom's inputs, it may reduce the \emph{dimensional rank} of the
post-compound variable set.  By the Buckingham $\pi$
theorem~\citep{Buckingham1914}, the resulting
function factors as
\begin{equation}
  f = x_{r_1}^{b_1}\!\cdots x_{r_p}^{b_p}\;\cdot\;g(\pi_1,\ldots,\pi_m) \, ,
  \label{eq:buckingham-factor}
\end{equation}
where the $x_{r_j}$ are the remaining (non-compound) variables and the
$\pi_k$ are dimensionless groups.  The monomial prefix carries the
physical dimensions of the output~$y$.

Let $\mathcal{D}_{\text{after}}$ be the set of dimension vectors of the
post-compound inputs (i.e.\ $[z]$ together with any extra variables),
and let $r = \operatorname{rank}\mathcal{D}_{\text{after}}$.  The
monomial prefix in Eq.~\eqref{eq:buckingham-factor} can only produce
dimensions in $\operatorname{span}\mathcal{D}_{\text{after}}$.
Therefore the compound is admissible if and only if
\begin{equation}
  [y_\varphi] \;\in\; \operatorname{span}\mathcal{D}_{\text{after}} \, ,
  \label{eq:rank-gate}
\end{equation}
where $[y_\varphi]$ is the dimension of the model target after any
$y$-transform~$\varphi$ (identity, $\log$, square, etc.).  In practice this
is tested by checking whether augmenting $\mathcal{D}_{\text{after}}$ with
$[y_\varphi]$ increases the rank.  If
$\operatorname{rank}(\mathcal{D}_{\text{after}} \cup \{[y_\varphi]\}) = r$,
the target is reachable and the compound is allowed.  Otherwise it is
rejected.

A rank drop is harmless whenever the lost dimensional directions are not
needed by the output (for a dimensionless target, $[y_\varphi]=\vect{0}$
lies in every subspace).

\subsection{Global Constraint Propagation (Buckingham--Sudoku, Level~2)}
\label{sec:units:global}

Local checks can approve a configuration that is \emph{globally} infeasible.
Consider a multiplicative decomposition
$\text{AST} = \text{NN}[x_0] \times (\text{NN}[x_1,x_3] \times
\text{NN}[x_2,x_4])$ for a problem where $[x_1]=[x_3]=\mathsf{L}$.  A
compound proposal $z = x_1/x_3$ is dimensionless, and the local Buckingham
check passes (the compound preserves the number of dimensionless degrees of
freedom).  However, once the inner atom sees only a dimensionless input, it
can only produce dimensionless output under the span semantics.  The outer
multiplication then cannot reach the required target dimension, a dead end
that Level~1 discovers only after expensive least-squares fitting.

Level~2 detects such dead ends \emph{before any fitting} by solving the
full dimensional constraint system globally.  The key insight is that
\emph{all dimensional constraints are linear over~$\mathbb{Q}$}, so the
entire system can be solved exactly in a single pass.

\paragraph{Feasible dimension subspaces.}
\label{sec:units:global:dimsubspace}

The situation is analogous to Sudoku.  Each node in the expression tree is a
cell whose ``value'' is a dimension vector, the known dimensions of inputs and
the target are the given clues, and the operator rules propagate constraints
until every cell is either uniquely determined or shown to be infeasible.

Concretely, each AST node is assigned an affine subspace of dimension-space
representing all dimensions the node \emph{could} take,
\begin{equation}
\mathcal{S} = \bigl\{\,\vect{d}_0 + \textstyle\sum_i c_i\,\vect{b}_i
\;\big|\; c_i \in \mathbb{Q}\,\bigr\} \, ,
\label{eq:dimsubspace}
\end{equation}
where $\vect{d}_0\in\mathbb{Q}^B$ is an offset (particular solution) and
$\{\vect{b}_i\}$ is a basis for the free directions.  Three special states
arise naturally: (i)~\emph{pinned} (basis empty), meaning the node's
dimension is uniquely determined; (ii)~\emph{unconstrained}, meaning all
dimensions are possible; and (iii)~\emph{empty} (system inconsistent),
meaning no dimension is feasible, i.e.\ a dead end.

Each AST node type imposes a linear constraint relating its dimension to
those of its children. This is applied both bottom-up (what a subtree can produce)
and top-down (what a child must produce).  

\begin{figure*}[t]
\centering
\includegraphics[width=\textwidth]{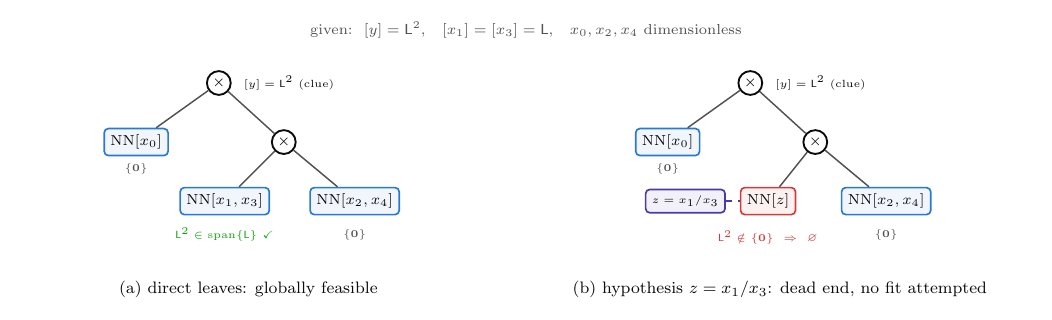}
\caption{Global dimensional feasibility (``Buckingham--Sudoku'',
\S\ref{sec:units:global}) on the worked example.  (a)~With direct
leaves the constraint system is solvable, and $\mathrm{NN}[x_1,x_3]$ can reach
the target dimension $\mathsf{L}^2$.  (b)~The compound $z=x_1/x_3$ passes
the local $\Pi$-gate (it is dimensionless), but it pins $\mathrm{NN}[z]$
to dimensionless output.  The required and achievable subspaces then have
empty intersection, and the hypothesis is rejected before any fitting.}
\label{fig:buckingham}
\end{figure*}

\paragraph{Algorithm: solve via RREF in one pass.}
\label{sec:units:global:algorithm}

Because every constraint is linear over~$\mathbb{Q}$, the entire
``Sudoku game'' reduces to a single linear system.
A symbolic traversal of the expression tree assigns each node a dimension
expression,
\begin{equation}
[\text{node}] = \vect{c}_0 + \sum_i \alpha_i\,\vect{U}_i \, ,
\label{eq:dimexpr}
\end{equation}
where $\vect{c}_0$ is a known constant part and each $\vect{U}_i$ is an
unknown dimension vector (one per unconstrained NN atom).  Equality
constraints from addition nodes, the root target, and the requirement that
transcendental arguments be dimensionless form a system
$\mat{A}\,\vect{x} = \vect{b}$ over~$\mathbb{Q}$.

Solving via reduced row echelon form (RREF) yields
\begin{equation}
\vect{x} = \vect{x}_0 + \sum_j t_j\,\vect{n}_j \, ,
\qquad t_j \in \mathbb{Q} \, ,
\end{equation}
where $\vect{x}_0$ is a particular solution and $\{\vect{n}_j\}$ spans
$\ker(\mat{A})$.  Each node's feasible dimension subspace
[Eq.~\eqref{eq:dimsubspace}] is then obtained by substituting this solution into
Eq.~\eqref{eq:dimexpr} and projecting.  The offset is obtained by evaluating the
node's dimension expression at $\vect{x}_0$, and the basis directions are
obtained by evaluating its linear part at each null-space vector
$\vect{n}_j$ and reducing to an independent set.

If RREF reveals an inconsistency ($0 = c \neq 0$ in any row), the entire
AST is infeasible, a dead end detected without any fitting.  The
computation is exact (rational arithmetic) and takes $\mathcal{O}(N A^2 B^3)$
time for $N$ nodes, $B$ base dimensions, and $A$ unknown atom dimensions
($A$ is at most a handful in practice), typically sub-millisecond for the
$N \leq 50$ trees encountered.

\subsection{Hypothesis Testing (Level~2)}
\label{sec:units:hypothesis}

Level~2 turns this solver into a cheap ``what-if'' test.  To ask whether an
atom may take a compound $z=x_i/x_j$, or split additively on a given
partition, we build the modified candidate AST, solve its global system from
scratch by RREF, and reject the proposal if any node's feasible subspace comes
out empty.  Figure~\ref{fig:buckingham} shows the canonical
case.  Level~1 thus acts as an $\mathcal{O}(1)$ pre-filter and Level~2 as a
global validator, both negligible against the least-squares fits
they save.  When dimensional analysis is active, factorized-search candidates
use a scaled parameterization separating unit-carrying scales from a
dimensionless core, so the resulting AST satisfies the global system by
construction.

\section{Benchmarks and Results}
\label{sec:benchmarks}

We evaluate NestyNet-SR in four settings.  The first is the full pipeline
(without the factorized symbolic search) on the AI~Feynman benchmark,
both noiseless and with additive noise (\S\ref{sec:full}).
The second is standalone factorized symbolic search, which consumes GS
carriers, on the same benchmark in oracle mode
(\S\ref{sec:aif-benchmark}).  The third is an
astrophysical theory-discovery example that recovers the galactic
tidal-radius law together with its anisotropic tidal invariant,
summarized in \S\ref{sec:jacobi_vignette} and distributed with the
code as a worked example.  The fourth is the real-data vignette, the
blind recovery of the baryonic acceleration coordinate from SPARC
galaxies against a deliberately verified target
(\S\ref{sec:sparc_vignette}).

\subsection{Full Pipeline (without Factorized Symbolic Search)}
\label{sec:full}

We now evaluate the NestyNet-SR pipeline in the noiseless regime
on all 120 AI~Feynman equations, but without the
factorized symbolic search.  The setup uses
the SRBench standardization~\citep{Udrescu2020,LaCava2021}, but with each
equation only sampled with $2\times10^3$  training and $2\times10^3$ validation 
points (instead of using the $10^5$ points allowed by the benchmark) drawn 
uniformly from the published domain, with no added noise.  We declare success 
when the discovered expression is algebraically identical to the ground-truth
expression.  For targets whose $|y|$ spans more than four decades and whose
identity-space fit fails or stalls, the surrogate fit is automatically retested under 
an $\operatorname{asinh}$ transform of the output.  

The pipeline recovers all $120$ equations exactly, at every input
dimensionality from one to seven-plus variables.  To the best of our knowledge, exact symbolic
recovery of all $120$ equations has not previously been reported on the
SRBench-standardized AI~Feynman
benchmark~\citep{Udrescu2020,LaCava2021,Tenachi2023PhySO}, so we regard this
as a new benchmark for exact recovery in this setting.  The full pipeline is also
relatively economical.  Running on a single CPU core (one thread), it solves each equation
in a median wall time of $6$~minutes, with $90\%$ of the 120 equations
recovered within $1.1$~hours.

For each problem the search is driven only by the sampled input--output
data, the variable ranges and names, and the dimension vectors of the input
and output channels, supplied as a units-only manifest released separately
from the answer key.  These metadata constrain which candidate expressions
are dimensionally admissible, but do not single out the target form among
the admissible ones.  The target expressions themselves never enter the
search.  The runs can be reproduced in a blinded mode in which the search
process has no file-system access to any file containing a target
expression, with algebraic equivalence scored afterwards by a separate
program, and blinded and non-blinded runs produce identical discovered
expressions.  The configuration is fixed: the benchmark driver issues the
same command line for every problem, and the main settings are collected in
Table~\ref{tab:hyperparams}.

\begin{table}[t]
\centering
\caption{Fixed hyperparameter configuration used identically for all 120
AI~Feynman equations (and, unless noted, the noisy runs and the
tidal-radius example).  Values are the pipeline defaults, and the benchmark
driver overrides none of them per problem.}
\label{tab:hyperparams}
\begin{ruledtabular}
\footnotesize
\begin{tabular}{l l}
\pcl{0.30\columnwidth}{Component} & \pc{0.60\columnwidth}{Setting} \\
\hline
\pcl{0.30\columnwidth}{Surrogate} & \pc{0.60\columnwidth}{dual-layer NestyNet, $f^{(1)}\!\circ f^{(0)}$, intermediate dim $M=n+2$, float64, $16$--$48$ segments per layer (progressive growth)} \\
\pcl{0.30\columnwidth}{Surrogate optimizer} & \pc{0.60\columnwidth}{Levenberg--Marquardt (direct solve), $\le 2\,500$ epochs, loss target $10^{-7}$, canonical initialization and segment-prior (evidence) mode on} \\
\pcl{0.30\columnwidth}{Data} & \pc{0.60\columnwidth}{$2\times10^3$ training / $2\times10^3$ validation points, uniform over the published domain, batch size $2\times10^3$} \\
\pcl{0.30\columnwidth}{Stage~A separability} & \pc{0.60\columnwidth}{mixed-second-derivative test tolerance $10^{-3}$, monomial rank-1 gate $\sigma_2/\sigma_1\le 0.1$ with confidence gate $0.85$} \\
\pcl{0.30\columnwidth}{Stage~A$\leftrightarrow$B loop} & \pc{0.60\columnwidth}{up to $5$ feedback iterations} \\
\pcl{0.30\columnwidth}{Stage~B} & \pc{0.60\columnwidth}{$\le 30$ outer iterations, $\le 2\,000$ LM epochs per candidate, $3$ backtracks, dynamic candidate policy, factorized symbolic search disabled for this table} \\
\end{tabular}
\end{ruledtabular}
\end{table}

To assess robustness, we repeat the benchmark with additive
Gaussian noise injected at several signal-to-noise ratios, as set by the SRBench rules.
The results are listed in Table~\ref{tab:full-noisy}.  Noise degrades recovery
under every selection scheme, yet close to half of the problems are successful
at the extreme $10\%$ level.

In Table~\ref{tab:full-noisy} we see that statistical selection (\S\ref{sec:stat_selection}) yields lower recovery 
than the heuristic acceptance (\S\ref{sec:stageB_accept}) at every noise level, as expected of a method that 
declines any candidate it cannot defend on sealed data against a simpler rival.  
Providing a selection committee (\S\ref{sec:coe}) generally overtakes the heuristic because it
widens the proposal pool.  Each scout and witness slice keeps the
$2\,000$-point training and validation sizes per fit, but the committee
draws on eight scout and eleven witness slices, disjoint from the
reference run and from each other, and five to eight times the computation of a
single-slice run.  Its statistical decisions use a separate sealed audit
partition of $20\,000$ rows, and only at $10^{-2}$ noise do the additional proposals fail
to survive certification.

\begin{table}[t]
\centering
\caption{Symbolic recovery on the AI~Feynman benchmark under additive
Gaussian noise, with factorized symbolic search disabled.  A problem is considered 
solved when the expression matches the target algebraically up to any
fitted constants, verified by refitting on noiseless data (tolerance $10^{-10}$).  
Round parentheses denote the heuristic acceptance of \S\ref{sec:stageB_accept}, 
unmarked values correspond to the statistical selection of \S\ref{sec:stat_selection}, and square
brackets add the committee of \S\ref{sec:coe}.}
\label{tab:full-noisy}
\begin{ruledtabular}
\begin{tabular}{ccc}
Noise level (${{\sigma}\over{\sigma_y}}$) & Solved / 120 & Rate \\
\hline
$0$ & (120) 120 [120] & (100\%) 100\% [100\%] \\
$10^{-3}$ & (105) 103 [110] & (87.5\%) 85.8\% [91.7\%] \\
$10^{-2}$ & (94) 91 [89] & (78.3\%) 75.8\% [74.2\%] \\
$10^{-1}$ & (56) 54 [62] & (46.7\%) 45.0\% [51.7\%] \\
\end{tabular}
\end{ruledtabular}
\end{table}

\subsection{AI Feynman Symbolic Regression Benchmark (Factorized Search, Oracle Mode)}
\label{sec:aif-benchmark}

To isolate the performance of the factorized symbolic search (FSS) method
from the full pipeline, we run it on the AI~Feynman benchmark, but restricted 
to the 115 equations with $n_{\mathrm{var}}\le6$.  The test is an 
\emph{oracle benchmark} without neural surrogate training, to remove
that source of uncertainty.  The search receives exact function values and dimensional 
metadata (512 fit points and a 2\,048-point search probe), and runs for 1\,400 mutation
iterations.  FSS evaluates exact target gradients solely to propose GS coordinates 
as carrier seeds.  A problem is considered solved using the structural
criterion discussed for Table~\ref{tab:full-noisy}.

This ablation removes almost all of the pipeline, including Stage~A separability, 
the Stage~B rewrite rules, and the surrogate-training
front end.  Because the fitted relation is $y=g(z)$, the
dimensions of a carrier $z$ need not equal those of $y$ before the outer map
$g$ is known (for example, a square-root map takes $L^2$ to $L$).  The
units check on such a carrier is therefore deferred: the carrier is
admitted with its internal units certified, and the carrier-to-target
relation is validated once the outer map has been fitted.  The result is
accordingly an exact-oracle-gradient numerical ablation.

\begin{table*}[t]
\centering
\caption{AI~Feynman benchmark: solve rate by number of input variables
($n_{\mathrm{var}}$).  \emph{Full pipeline} is the complete
Stage~A/Stage~B stack run on all 120 noiseless equations with factorized
symbolic search disabled.  The oracle-mode FSS benchmark is run on the 115
equations with $n_{\mathrm{var}}\le6$.  The counts are structural recoveries under the
criterion of Table~\ref{tab:full-noisy}. FSS consumes exact-oracle-gradient GS carrier proposals, with carrier
units validated after the outer-map fit as described in the text.}
\label{tab:aif-benchmark}
\begin{ruledtabular}
\footnotesize
\begin{tabular}{lccccccccc}
Method & $n_{\mathrm{var}}=1$ & 2 & 3 & 4 & 5 & 6 & 7+ & All & Rate \\
\hline
Full pipeline & 1/1 & 15/15 & 37/37 & 32/32 & 20/20 & 10/10 & 5/5 & 120/120 & 100\% \\
\textbf{FSS} & \textbf{1/1} & \textbf{15/15} & \textbf{34/37} & \textbf{24/32} & \textbf{13/20} & \textbf{1/10} & \textbf{--} & \textbf{88/115} & \textbf{76.5\%} \\
\end{tabular}
\end{ruledtabular}
\end{table*}

Table~\ref{tab:aif-benchmark} summarizes the results.  FSS recovers 88 of
the 115 eligible equations (76.5\,\%).  For instance, for AI~Feynman problem \#3, GS supplies
$z=(x_0-x_1)^2+(x_2-x_3)^2$, after which the existing power outer map fits
an exponent of $0.5$.  Its search-probe MSE is
$2.13\times10^{-31}$ and its independent dense-holdout MSE is
$2.47\times10^{-31}$, with a runtime of 69.4~s.

Across all 115 equations, FSS has median and mean wall times of 30.0~s and
553~s (hard and unsolved mutation-search cases dominate the budget).  
The remaining unvalidated equations are generally deeply
nested five- and six-variable expressions with non-monomial carriers or
additive inner structure that resists the current peel decomposition.
Note that these results are from a one-shot pass, i.e., the factorized symbolic-search method must
solve the problem immediately in its entirety, without recourse to any of the
other simplification methods in NestyNet-SR's extensive arsenal.

\subsection{Galactic Tidal-Radius Example}
\label{sec:jacobi_vignette}

Before turning to the sky we validated the discovery layers end to end on
a controlled theory-recovery study, the circular-orbit Jacobi (tidal)
radius $r_J=\left[\mu/(4\Omega^{2}-\kappa^{2})\right]^{1/3}$ of star
clusters in an ensemble of mock galactic hosts, where
$\mu=GM_{\rm cl}$ for cluster mass $M_{\rm cl}$, and $\Omega$ and
$\kappa$ are the circular and epicyclic frequencies of the cluster's
orbit about its host.  The generalized-symmetry layer
discovers the anisotropic invariant $4\Omega^{2}-\kappa^{2}$, and the pipeline assembles
the exact law in closed form, unifying the classical Hill/Roche and
flat-rotation-curve limits including the cube-root mass exponent.  The
complete study, including the support-geometry design on which
the recovery depends and the measured gradient-noise ladder (symmetry
discovery inherits the surrogate's derivative quality, certifying only
while gradients are accurate to ${\sim}10^{-3}$), is
distributed with the code as a worked example.\footnote{In the directory
\texttt{examples/jacobi\_tidal} of the NestyNet-SR repository.}

\subsection{Blind Recovery of the Baryonic Acceleration Coordinate from SPARC}
\label{sec:sparc_vignette}

\begin{figure*}[t]
\centering
\includegraphics[width=0.92\textwidth]{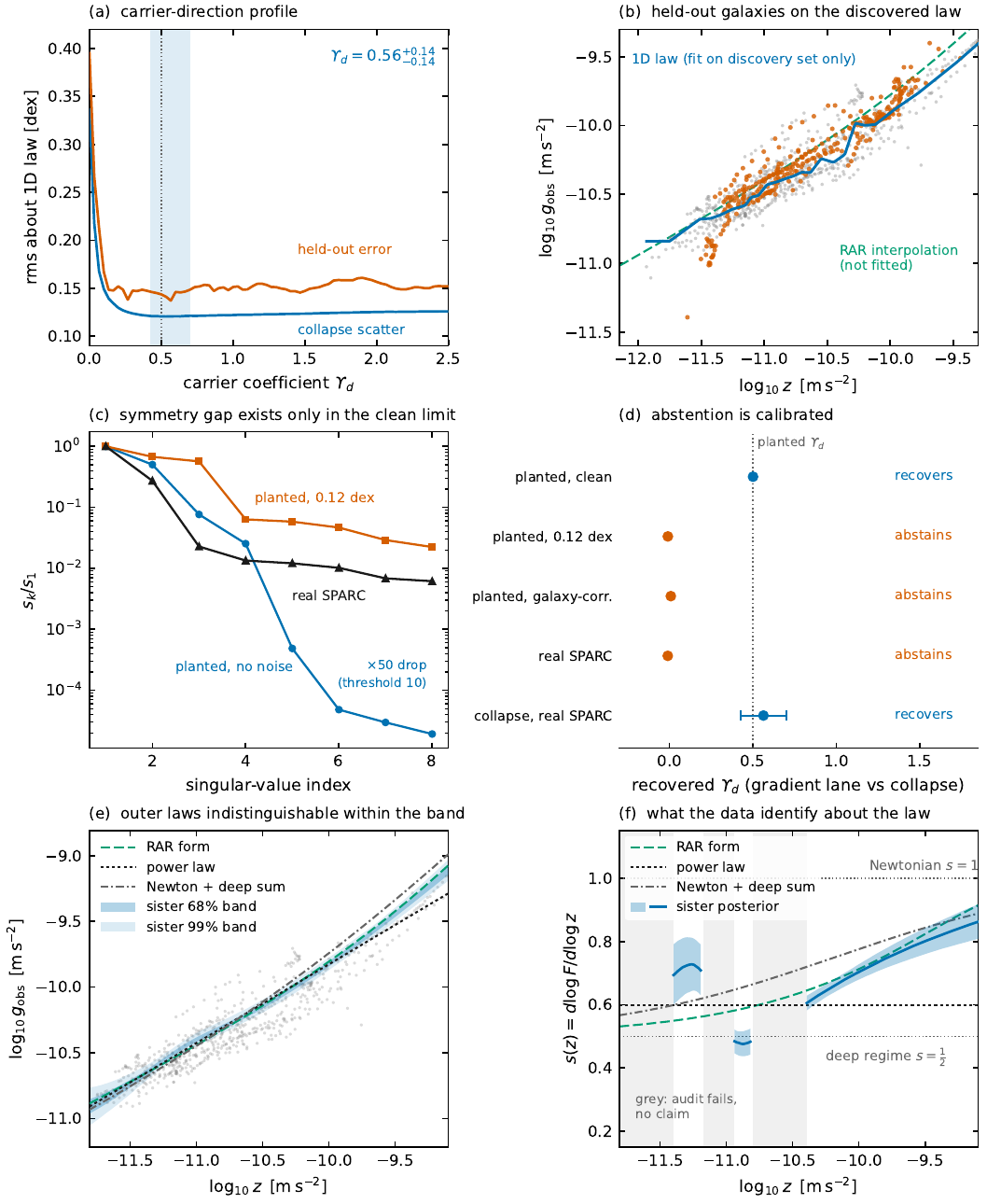}
\caption{The SPARC vignette (gold sample, whole-galaxy
discovery/held-out split, 34/14).
(a)~Collapse scatter and held-out error select the same carrier
direction, $\Upsilon_{d}=0.56^{+0.14}_{-0.14}$ (band, whole-galaxy
bootstrap).  (b)~The law fitted on discovery galaxies predicts the
held-out galaxies (orange) at 0.13 dex; dashed, the RAR interpolation, not fitted.
(c)~Determining-operator spectrum.  For the planted noise-free
carrier the smallest singular value drops a factor of 50 below the
rest, a numerically exact symmetry (certification requires a factor of
10); 0.12 dex of scatter (the observed scatter about the law) erases the drop and the operator abstains.
(d)~Recovered $\Upsilon_{d}$: the gradient certificate succeeds only
in the clean limit; the collapse recovers it from real data.  (e)~The sister 
model 68\% and 99\% bands, with galaxies as the independent units, and the
candidate outer laws threaded through them, indistinguishable on
held-out galaxies.  (f)~The slope posterior $s(z)$ on its certified
domain (gray, no claim), rising from the deep-regime $\tfrac12$
toward the Newtonian $s=1$.}
\label{fig:sparc_vignette}
\end{figure*}

As a real-data demonstration we ask whether the system, given only 
separate gas and stellar-disk contributions to galaxy rotation curves,
discovers that disk dynamics scale via a single baryonic
acceleration coordinate.  The SPARC database \citep{Lelli2016} provides
$V_{\rm obs}(R)$ together with the gas and disk contributions, with the
disk at a nominal Spitzer $3.6\,\mu$m mass-to-light ratio of unity.  The radial acceleration relation
\citep[RAR;][]{McGaugh2016,Lelli2017} is conventionally studied by
forming the baryonic acceleration $g_{\rm bar}$ \emph{in advance} and
fitting the one-dimensional relation $g_{\rm obs}(g_{\rm bar})$, a
problem to which exhaustive symbolic regression has been applied
\citep{Desmond2023}.  Here we take the aggregation itself as unknown.  The
inputs are $\mathbf{x}=(g_{\rm gas},\,g_{\rm disk}^{(\Upsilon=1)})$
with $g_{\rm comp}=V_{\rm comp}|V_{\rm comp}|/R$, the target is
$g_{\rm obs}=V_{\rm obs}^{2}/R$, and the sought structure is
\begin{equation}
z=g_{\rm gas}+\Upsilon_{d}\,g_{\rm disk}^{(\Upsilon=1)} \, ,
\qquad
g_{\rm obs}=F(z) \, ,
\label{eq:sparc-carrier}
\end{equation}
where the gas term fixes the scale gauge and $\Upsilon_{d}$ is 
the stellar disk mass-to-light ratio in solar units at $3.6\,\mu$m.  
Our choice of scientific problem is deliberate:
the $\Upsilon_{d}$ scale, geometry sensitivities, and form identifiability
are externally established, so the literature counterparts must be 
reproduced blind, and our contribution is the hands-off protocol itself.

We first make a selection for kinematic regularity, taking galaxies with
quality $Q<3$, $e_{V}/V<0.1$, no bulge, inclination $40\arcdeg\le i\le75\arcdeg$,
types Sbc--Sm, and containing at least eight radial measurements.  
This leaves a gold sample of 48 galaxies with 945 points, of which 
34 galaxies form the discovery set and 14 are held out entirely.

The carrier is identified by the class-shared collapse, the
multi-dataset structure of \S\ref{sec:class_param_sr} specialized to
Eq.~\eqref{eq:sparc-carrier}, with $F$ nonparametric,
and each galaxy being a separate experiment.  Profiling over the carrier direction
with (whole galaxy) bootstraps gives
$\Upsilon_{d}=0.56^{+0.14}_{-0.14}$, consistent with the
population synthesis expectation of ${\simeq}0.5$ at $3.6\,\mu$m
adopted by \citet{Lelli2016}, and the fully held-out prediction
error selects the same direction, with the trivial gas-only carrier
($\Upsilon_{d}=0$) rejected at 0.34--0.40 dex
(Fig.~\ref{fig:sparc_vignette}a).  Generalization provides the decisive test.
A one-dimensional law fitted through $z$ on the
discovery-set galaxies predicts the held-out galaxies to within an rms
scatter of 0.13 dex in $\log g_{\rm obs}$, while an unrestricted two-dimensional surrogate trained on the same rows
predicts them at 0.65 dex (Fig.~\ref{fig:sparc_vignette}b), a grouped
counterpart of the fundamentality analyses of \citet{Stiskalek2023}.
The factorized search, run blind on the two components, corroborates
this: its leading candidates are the small-rational renderings
$g_{\rm gas}+g_{\rm disk}/3$ and $g_{\rm gas}+g_{\rm disk}/2$ at the
same held-out level, with non-carrier candidates trailing beyond 0.15
dex.

Each galaxy is a thin locus in component-acceleration space, so the
transverse derivative is constrained only across galaxies, and at the
level of scatter in the catalog (0.12 dex) the surrogate's transverse gradients are set
by its smoothness.  A planted-carrier control makes this
quantitative.  Keeping the real input rows but replacing the observed
accelerations with a synthetic target, a known law over a planted
coordinate (the RAR interpolating form with $\Upsilon_{d}=0.5$), the
determining operator recovers the planted translation
algebra of Eq.~\eqref{eq:sparc-carrier}, its smallest singular value
falling a factor of 50 below the rest of the spectrum (certification
requires a factor of 10), and the gradient readout returns
$\Upsilon_{d}=0.500$ exactly, while 0.12 dex of scatter,
row-independent or galaxy-correlated, closes that separation and the
operator abstains on data that contain the carrier by construction
(Fig.~\ref{fig:sparc_vignette}c,d).  The abstention is therefore
calibrated, discovery requires cross-galaxy information, and on
the clean control the factorized search closes the loop, recovering
the carrier as $2\,g_{\rm gas}+g_{\rm disk}^{(\Upsilon=1)}$ and the
planted law at $3\times10^{-4}$ dex.

With the coordinate fixed, candidate outer laws compete on the
held-out galaxies.  The RAR interpolating function of
\citet{McGaugh2016}, the simple interpolation function, the
superposition $F=z+\sqrt{g_{\dagger}z}$, a pure power law, and a
log-parabola all land within 0.106--0.116 dex of error-weighted
held-out scatter, against a galaxy-bootstrap uncertainty of 0.022 dex.
None improves on the nonparametric collapse law itself, 0.118 dex on
the same metric, by more than that uncertainty.  The common acceleration scale
is $g_{\dagger}\simeq0.96\times10^{-10}\,\mss$
(Fig.~\ref{fig:sparc_vignette}e), slightly below the $1.20\times10^{-10}\,\mss$ of \citet{Lelli2017}.
With the free factorized-search
slate included, the complexity--accuracy Pareto front is flat within
that uncertainty: no candidate can be isolated, and added structure yields no
accuracy improvement on held-out samples.  We therefore report the outer law
as a family, in agreement with \citet{Desmond2023}, and promote no
unique interpolation function.  Instead, what the data do identify is delivered
through the statistical layer of Paper~II \citep{NestyNet2026b} (the ``sister'' model).
Conditional on the discovered carrier, with $\Upsilon_{d}$ and the
catalog geometry held fixed, coherent sister model draws of the law and its derivative,
with galaxies rather than individual rotation-curve points as the independent
units (a sandwich covariance that discounts only the correlated component of
the residuals within each galaxy), give the posterior
of the local slope $s(z)=d\log F/d\log z$ on the domain where its
certificates hold.  Across the main certified range,
$4\times10^{-11}$ to $8\times10^{-10}\,\mss$, $s$ rises from $0.61\pm0.02$
through $0.70\pm0.02$ at $10^{-10}\,\mss$ to $0.86\pm0.06$, and on the
certified island at $1.2\times10^{-11}\,\mss$ it is $0.48\pm0.04$
(Fig.~\ref{fig:sparc_vignette}f).  The deep-regime slope of
$\tfrac12$, from which the baryonic Tully--Fisher scaling follows, is
recovered as a measurement.

Distance and inclination move whole galaxies in the acceleration
plane, $g_{\rm obs}\propto D^{-1}\sin^{-2}i$ with the components
nearly distance-independent \citep{Li2018}.  Refitting with the distance $D_k$ and
inclination $i_k$ of each galaxy $k$ as free parameters under Gaussian priors from the catalog
uncertainties, with the sample-coherent combination constrained to
zero mean, reduces
the weighted scatter from 0.112 to 0.065 dex, comparable to the 0.057
dex of \citet{Li2018}, with unit-variance nuisance pulls.
$\Upsilon_{d}$ shifts by $+0.25$, well inside the 0.70 full width of the 68\% 
galaxy-bootstrap interval of the released fit, which is wider than the 
fixed-geometry $\pm0.14$ because the freed geometry absorbs part of the constraint.
The discovered coordinate is thus robust to the uncertain galaxy
geometry, but an audit against resolved velocity fields
\citep{Trachternach2008} is left to future work.

Thus every conclusion with a literature counterpart is
reproduced blind, and the protocol adds the
aggregation question, the held-out scoring, the calibrated abstention,
and the certified slope posterior.  The law becomes simple only in the
discovered coordinate, and every claim is scoped to the domain on
which its certificate holds.

\section{Conclusions}
\label{sec:conclusions}

The present paper advances symbolic regression by providing derivative-accurate 
decomposition, and by implementing a new symbolic search method.

\emph{Derivative-accurate decomposition engine.}  All symbolic expressions
are represented as PyTorch-integrated abstract syntax trees with compound
variable support, atom tagging for module reuse, and a comprehensive
leaf-type library.  We use the accurate gradients and Hessians from NestyNet
models to make classical decomposition motifs substantially more reliable,
while also extending them to approximate, overlapping,
offset-multiplicative, and richer compound-coordinate structure.  In the
current AI~Feynman setting this structured simplification stack already
suffices to attain exact symbolic recovery. This appears to be the first
such result reported in the literature.

\emph{New symbolic-search method.}  The second contribution is a standalone
factorized symbolic-search algorithm organized around a skeleton--mapping
factorization, residual basins, archive-guided UCB mutation and repair,
closure-native basis proposals, optional continuous skeleton refinement,
dimensional filtering, and targeted peel-presearch for difficult
compositional families.  On the AI~Feynman benchmark in oracle mode, FSS
consumes exact-gradient GS carrier proposals and yields 88 recoveries among
the 115 eligible equations (76.5\,\%), in a deliberately hard one-shot
challenge.  That benchmark forgoes the Stage~A and Stage~B simplifications
available in the pipeline, as well as problem-specific equation templates
and the reinforcement learning and genetic programming on which other SR
pipelines are built.  We therefore view this factorized symbolic search as a
genuine algorithmic result in its own right.  Within the full NestyNet-SR
pipeline, it serves as the deliberate escalation layer when decomposition
and cheaper targeted rewrites no longer close the problem cleanly.  Its main
power over the rewrite tier is that it can form arbitrary compositions of
functions while carrying their nonlinear parameters, so constants that many
direct tree-search methods generate via tokens or optimize one finished
candidate expression at a time are absorbed into the scoring of whole
families at once.  Note that the AI Feynman benchmark does not provide a
favorable showcase of the method, as it contains virtually no parameters
inside nonlinear operators to discover.  We expect FSS to shine over
competitors in more realistic scenarios.

Factorized symbolic search is also fast.
Because the factorization lets a single carrier stand in for a whole family
of calibrated renderings, and because dimensional filtering prunes the
enumeration aggressively, nearly half of the AI~Feynman targets are solved
by the deterministic presearch in under ten seconds, and the median wall
time is 30~s on one CPU core (\S\ref{sec:aif-benchmark}).  Computational
cost is a central practical bottleneck for symbolic regression, so this
efficiency is itself a significant advantage.  More broadly, we think the
right lesson is not that generic search should replace structured rewrites,
but that a strong symbolic regression engine should be tiered, by exploiting
the cheapest high-confidence scientific motifs first, then escalating to
broader factorized symbolic search only when the evidence demands it.

Two further components support the pipeline: the dimensional analysis of
\S\ref{sec:units}, enforced both locally and by the global
``Buckingham--Sudoku'' solver, and the post-Stage-B multi-dataset refinement
of \S\ref{sec:class_param_sr}, which separates class from experiment
parameters and recovers derived invariants.

A third layer, the generalized-symmetry operator of \S\ref{sec:gs},
recovers the Stage~A coordinate families from one determining equation and
reaches coordinates outside their vocabulary.  The same philosophy extends to
differential-equation discovery (developed in Paper~IV), where on-shell
generator recovery, relative-invariance certificates, and symmetry reductions
seed the search, and where the generator family broadens beyond the affine
case to quadratic point symmetries and to noncanonical Poisson geometry,
including its Casimir invariants.

The galactic tidal-radius example (\S\ref{sec:jacobi_vignette}) demonstrates
that the algorithm can go beyond recovering individual formulas to exposing
the meta-structure that links them.  Symbolic regression across an ensemble
of physical regimes can expose the invariant underlying a family of
empirical laws, not just a single static closed-form relation.

The SPARC vignette (\S\ref{sec:sparc_vignette}) extends this demonstration
to the sky focussing on a research project that has already been independently
explored.  A blind run reproduces the
established results of the radial acceleration relation while adding the
aggregation question, held-out-galaxy scoring, planted-truth abstention
calibration, and a certified slope posterior, which established analyses,
with $g_{\rm bar}$ formed in advance, did not pose.


The Stage~B rewrite rules are heuristic in nature and may miss optimal
analytical forms.  Factorized symbolic search substantially broadens the
search space, but the cases it misses in the AI~Feynman benchmark show that
deeply nested high-arity expressions remain challenging for such an
approach.  Computational cost also scales exponentially with the number of
variables for exhaustive separability search, which limits the approach to
moderate dimensionality without further pruning strategies.  More systematic
use of fixed-AST joint fitting, richer multi-dataset invariants, and
stronger guidance for the mutation phase of factorized symbolic search are
natural next steps, as are broader benchmark suites beyond
AI~Feynman.  Better criteria for when to escalate from targeted
rewrites to the more general search layer are also an obvious next target
for the live Stage~B controller.

\section*{Data Availability}

The codebase is available at \url{https://github.com/RodrigoIbata/NestyNet_SR}.
The benchmark input data are archived at
\dataset[doi:10.5281/zenodo.21390410]{https://doi.org/10.5281/zenodo.21390410}.
The benchmark evidence and reproducibility package can be found at
\dataset[doi:10.5281/zenodo.22027264]{https://doi.org/10.5281/zenodo.22027264}, 
which contains the benchmark evidence capsules, and the FSS audit.  The Jacobi 
and SPARC vignette scripts and result summaries are distributed with the codebase.

\begin{acknowledgments}
RI gratefully acknowledges funding in the initial stages of this project from the European Research Council (ERC) under the European Union's Horizon 2020 research and innovation program (grant agreement No. 834148). We gratefully acknowledge the High Performance Computing center of the Universit\'e de Strasbourg for a very generous time allocation and for their support over the development of this project.
\end{acknowledgments}

\software{NestyNet \citep{NestyNet2026a},
PyTorch \citep{Paszke2019},
NumPy \citep{Harris2020},
SciPy \citep{Virtanen2020},
SymPy \citep{Meurer2017},
Matplotlib \citep{Hunter2007}.}

\bibliographystyle{aasjournalv7}
\bibliography{nestynet_paper3}

\end{document}